\documentclass[trackchanges,twocolumn,twocolappendix]{aastex7}

\usepackage[version=4]{mhchem}
\usepackage{graphicx}
\usepackage{subcaption}
\usepackage[margin=1in]{geometry}
\usepackage{soul}
\usepackage{tabularx}
\usepackage{multirow}%
\usepackage{booktabs} 
\usepackage{seqsplit}
\usepackage{threeparttable}
\usepackage{bm}
\usepackage{placeins}   % for \FloatBarrier
\usepackage{float}      % for [H] placement when you truly must fix a float here
\allowdisplaybreaks     % let long equation blocks split across pages

\begin{document}

\title{Coupled 1D Chemical Kinetic-Transport and 2D Hydrodynamic Modeling Supports \\a modest 1--1.5$\times$ Supersolar Oxygen Abundance in Jupiter’s Atmosphere}

\author[0000-0002-1551-2610]{Jeehyun Yang}
\affiliation{Jet Propulsion Laboratory, California Institute of Technology,
Pasadena, CA 91109, USA}
\affiliation{Division of Geological and Planetary Sciences, California Institute of Technology, Pasadena, CA 91125, USA}
\affiliation{Department of Astronomy and Astrophysics, The University of Chicago, Chicago, IL 60637, USA}
\email[show]{jeehyuny@uchicago.edu}  

\author[0000-0002-3350-4243]{Ali Hyder} 
\affiliation{Jet Propulsion Laboratory, California Institute of Technology,
Pasadena, CA 91109, USA}
\email{ali.hyder@jpl.nasa.gov}

\author[0000-0003-2215-8485]{Renyu Hu}
\affiliation{Jet Propulsion Laboratory, California Institute of Technology, Pasadena, CA 91109, USA}\affiliation{Department of Astronomy \& Astrophysics, The Pennsylvania State University, University Park, PA 16802, USA}
\affiliation{Center for Exoplanets and Habitable Worlds, The Pennsylvania State University, University Park, PA 16802, USA}
\affiliation{Institute for Computational and Data Science, The Pennsylvania State University, University Park, PA 16802, USA}
\email{rgh5611@psu.edu}

\author[0000-0003-2279-4131]{Jonathan I. Lunine}
\affiliation{Jet Propulsion Laboratory, California Institute of Technology,
Pasadena, CA 91109, USA}
\affiliation{Division of Geological and Planetary Sciences, California Institute of Technology, Pasadena, CA 91125, USA}
\email{jonathan.i.lunine@jpl.nasa.gov}

\begin{abstract}

Understanding the deep atmospheric composition of Jupiter provides critical constraints on its formation and the chemical evolution of the solar nebula. In this study, we combine one-dimensional thermochemical kinetic-transport modeling with two-dimensional hydrodynamic simulations to constrain Jupiter’s deep oxygen abundance using carbon monoxide (CO) as a proxy tracer. Leveraging a comprehensive chemical network generated by Reaction Mechanism Generator (RMG), we assess the impact of updated reaction rates, including the often-neglected but thermochemically significant Hidaka reaction (\ce{CH3OH} + H → \ce{CH3} + \ce{H2O}). Our 1D-2D coupled approach supports a modest supersolar oxygen enrichment of $1.0–1.5\times$ the solar value. We also present a method for deriving Jupiter's eddy diffusion coefficient ($K_{\rm zz} = 3 \times 10^6$ to $5 \times 10^7$ cm$^2$/s) from 2D hydrodynamic simulations using the quasi steady-state approach. This method is applicable to exoplanet atmospheres, where $K_{\rm zz}$ remains highly uncertain despite its strong influence on atmospheric chemistry. Finally, our results imply a significantly elevated planetary carbon-to-oxygen (C/O) ratio of $\sim$2.9, highlighting the importance of clarifying the mechanisms behind the preferential accretion of carbon-rich material during Jupiter's formation. 
By integrating thermochemical and hydrodynamic processes, our study offers a more complete framework for constraining chemical and dynamical processes in (exo)planetary atmospheres.

\end{abstract}

\keywords{\uat{Jupiter}{873}, \uat{Planet formation}{1241}, \uat{Planetary atmospheres}{1244}, \uat{Protoplanetary disks}{1300}}


\section{INTRODUCTION} \label{sec:introduction}

Understanding the composition of giant planets is key to constraining their formation mechanisms. In particular, the abundances of carbon and oxygen provide insight into where a planet formed within the protoplanetary disk, as well as its accretion history and subsequent evolution \citep{oberg2011effects, mousis2019jupiter}. For this reason, measuring species such as water, methane, and ammonia, and characterizing the chemistry of Jupiter’s deep atmosphere (as probed by the Galileo and Juno missions), has been a major focus in planetary science \citep{niemann1998composition, bezard2002carbon, wong2004updated, bjoraker2018gas}. Among these, determining the oxygen abundance in Jupiter’s deep atmosphere remains particularly challenging since water, the primary oxygen-bearing species, undergoes condensation and complex hydrodynamic processes, often resulting in locally depleted abundances and complicating efforts to measure a representative mixing ratio for the entire envelope. Instead, carbon monoxide, which is physically and chemically stable under Jupiter's deep atmospheric conditions and thus readily detectable, is frequently used as a proxy to constrain the deep oxygen abundance through one-dimensional (1D) chemical kinetic-transport models \citep{prinn1977carbon, visscher2010deep, wang2015new, Wang_2016_Jupiter, cavalie2023subsolar}.

Recently, \citet{cavalie2023subsolar} reported a subsolar oxygen abundance of 0.3 times the protosolar value using a 1D chemical kinetic-transport model that incorporated an updated chemical network from \cite{venot2020new} (hereafter referred to as V20) and assumed an eddy diffusion coefficient ($K_{\rm zz}$) of 10\textsuperscript{8} cm\textsuperscript{2}/s. The chemical network developed by \citet{venot2020new} (i.e., V20) excluded the reaction H + \ce{CH3OH} $\rightarrow$ \ce{CH3} + \ce{H2O} (hereafter referred to as the Hidaka reaction), originally derived by \citet{hidaka1989thermal}. While \cite{venot2020new} found that hot Jupiter chemistry is relatively insensitive to the inclusion of this reaction, they showed that it significantly affects CO/\ce{CH4} interconversion in cooler, higher-pressure environments such as warm Neptunes (for example, in GJ 436~b, including the Hidaka reaction led to a $\sim$2.5$\times$ increase in the quenched CO mole fraction), T dwarfs, and potentially Neptune and Uranus. This, then, raises the question of whether it is appropriate to exclude the Hidaka reaction in modeling Jupiter’s atmosphere \citep{cavalie2023subsolar}, given the relatively cool temperatures and high pressures present in its deep atmospheric layers.

In addition to limitations in the chemical network, the 1D kinetic-transport framework simplifies vertical mixing by representing it with a single eddy diffusion coefficient ($K_{\rm zz}$). Although several 1D models include vapor phase change behavior and condensible species, the underlying mixing strength is assumed to be decoupled from these effects and treated independently. In reality, increased condensation near the cloud layers can inhibit mixing by stabilizing the atmosphere against convective instability. As a result, the 1D model's ability to accurately constrain Jupiter's oxygen abundance is limited. To address this limitation, \citet{hyder2025supersolar} used a 2D hydrodynamic modeling framework to simulate atmospheric dynamics at and below the water cloud level. They incorporated simplified CO thermochemistry by using the temperature- and pressure-dependent chemical lifetime of CO to capture the influence of hydrodynamic transport on CO abundance in Jupiter’s troposphere. Their results support a supersolar enrichment of Jupiter’s deep atmosphere, with metallicities estimated between 2.5 and 5 times solar. However, the CO chemical timescale in \citet{hyder2025supersolar} was obtained by applying a scaling factor to the \ce{CO-CH4} interconversion timescale derived by \citet{wang2015new}, which itself was based on the chemical network of \citet{venot2012chemical} (hereafter V12). Given that the V12 network has since been substantially updated in \citet{venot2020new} (V20), it would be preferable to empirically derive the CO chemical timescale using the more recent and more comprehensive chemical network to precisely capture \ce{CO-CH4} interconversion and its coupling with hydrodynamic behavior.

Over the past decade, advances in computational chemical engineering, particularly through rate-based algorithms, have enabled more precise and comprehensive predictive chemistry \citep{Gao_2016, liu2021reaction, RMG-database}. This tool, also known as Reaction Mechanism Generator (RMG), assesses and determines key chemical species and reactions based on iterative simulations of the chemical system of interest, producing objective, user-independent, and more complete networks than those traditionally built by individuals. Notably, RMG has not only been widely used in chemical engineering but has also enabled successful simulations of exoplanet atmospheric chemistry recently by coupling its automated reaction mechanism generation with 1D photochemical kinetic-transport modeling using the \texttt{EPACRIS} framework \citep{yang2024automated}. This approach has been applied to planets such as WASP-39~b, WASP-80~b, K2-18~b, and TOI-270~d, yielding results that are consistent with the JWST observations\citep{yang2024automated, yang2024chemical, benneke2024jwst, hu2025water}.

Because \ce{CO-CH4} interconversion in the deep atmospheres of giant planets is primarily driven by thermochemistry and vertical mixing rather than photochemistry, applying a rate-based, RMG-generated chemical network in combination with a two-dimensional hydrodynamic transport model enables accurate characterization of these processes. This approach can help constrain the deep oxygen abundance of Jupiter, and at the same time, can be extended to Neptune and Uranus, whose internal compositions remain poorly understood. This is a well-recognized problem in planetary science. For example, the 2020 Planetary Science and Astrobiology Decadal Survey highlights the need for improved understanding of the formation and evolution of the ice giants \citep{decadalsurvey}.

Given this background, in this work we employ both a 2D hydrodynamic model to simulate atmospheric microphysics and vertical transport, and an automated, pressure- and temperature-dependent chemical network generator to capture complex thermochemical kinetics across a wide range of conditions. By leveraging these two state-of-the-art computational tools, we aim to simulate \ce{CO-CH4} interconversion in Jupiter’s deep troposphere with unprecedented detail and thereby place tighter constraints on its deep oxygen abundance, a key tracer of giant planet formation.

\section{METHODS} \label{sec:method}
\subsection{Oxygen and Other Elemental Parameterization of Jupiter's Tropospheres} \label{subsec:elemental}
To investigate the various oxygen abundance scenarios in Jupiter's troposphere, we first constrained the abundances of helium (He), carbon (C), nitrogen (N), and sulfur (S) using the values from Table 2 in \cite{Wang_2016_Jupiter}: [\ce{He}]/[\ce{H2}] = 0.157, and thus [\ce{He}]/[H] $\sim \tfrac{1}{2}$[\ce{He}]/[\ce{H2}] = $7.85 \times 10^{-2}$; [\ce{CH4}]/[\ce{H2}] = $2.37 \times 10^{-3}$, and thus [C]/[H] $\sim \tfrac{1}{2}$[\ce{CH4}]/[\ce{H2}] = $1.19 \times 10^{-3}$; [\ce{NH3}]/[\ce{H2}] = $6.64 \times 10^{-4}$, and thus [N]/[H] $\sim \tfrac{1}{2}$[\ce{NH3}]/[\ce{H2}] = $3.32 \times 10^{-4}$; [\ce{H2S}]/[\ce{H2}] = $8.90 \times 10^{-5}$, and thus [S]/[H] $\sim \tfrac{1}{2}$[\ce{H2S}]/[\ce{H2}] = $4.45 \times 10^{-5}$. These values are based on Galileo Probe Mass Spectrometer (GPMS) measurements of the major gas species’ mixing ratios relative to \ce{H2} \citep{niemann1998composition, wong2004updated}. Next, we adopted an astronomical log-scale abundance, \textit{A(E)}, using a value of 8.73 for oxygen, corresponding to the present-day solar oxygen abundance from Table 3 of \cite{lodders2021relative}. We set this oxygen abundance as O/H=1$\times$\textit{Z}\textsubscript{$\odot$}, where \textit{Z}\textsubscript{$\odot$} refers to the solar metallicity. For example, if the oxygen enrichment factor $E$ is 2.3, then the O/H ratio is given by $10^{8.73+\log_{10}2.3-12}\sim1.24\times10^{-3}$. Finally, we derived the elemental abundances by normalizing the sum of H, He, C, N, O, and S to unity, and the resulting elemental composition profiles for each O/H enrichment scenario are listed in Table A~\ref{tab:elemental_parameterization}. It has to be noted that since we adopted the carbon abundance constrained by the GPMS measurement, the resulting carbon-to-oxygen ratio (C/O) for the O/H=1$\times$\textit{Z}\textsubscript{$\odot$} scenario is 2.21, which is significantly higher than the present-day solar C/O = 0.55 \citep{lodders2021relative} (corresponding to $\sim$4$\times$ the solar C/O using values from \citet{lodders2021relative} and $\sim$4.4$\times$ using \citet{asplund2009chemical}). Within this framework, the \ce{H2O} abundance (2.5$\times10^3$ ppm) inferred for Jupiter's equatorial region by the Juno Microwave Radiometer (MWR) measurements in 2020 \citep{li2020water} corresponds to O/H = 2.7$^{+2.4}_{-1.7}\times$\textit{Z}\textsubscript{$\odot$} and C/O = 0.82$^{-0.39}_{+1.39}$, and the latest analysis of the O/H ratio from Juno data by \cite{LI2024116028} corresponds to O/H = 4.5$\pm3.1\times$\textit{Z}\textsubscript{$\odot$} (It has to be noted that \citet{LI2024116028} adopted the $A(E)$ value from \citet{asplund2009chemical}, and reported O/H=4.9$\pm3.4\times$\textit{Z}\textsubscript{$\odot$}).

\subsection{The T-P and K$_{zz}$ profiles for Jupiter's Tropospheres} \label{subsec:inputs}
The temperature–pressure ($T$–$P$) profile of Jupiter is adopted from \cite{seiff1998thermaljupiter, simon2006jupiter}. The vertical eddy diffusion coefficient ($K_{\mathrm{zz}}$) is taken from \cite{moses2005photochemistry}, which was developed to reproduce observed abundances of various atmospheric species \citep{atreya1981jupiter, vervack1995jupiter, yelle1996structure, lara1998high, edgington1999ammonia, drossart1999fluorescence, drossart2000methane, bezard2002carbon, lellouch2002origin, moreno2003long}. Since the major purpose of the current study is to investigate the quenching of carbon monoxide in Jupiter's deep atmosphere ($P \geq 10$~bar) in detail, we have varied the deep atmospheric $K_{\text{zz}}$ value from $5\times10^6$ to $10^9$~cm$^2$\,s$^{-1}$, with a nominal value of $10^8$~cm$^2$\,s$^{-1}$ (see black solid line in Figure~B\ref{fig:tp_kzz_profiles} in Appendix~\ref{sec:appendix_inputs}).

\subsection{Chemical Networks} \label{subsec:chemical_network}
The chemical network constructed using the Reaction Mechanism Generator \cite[RMG]{Gao_2016, RMG-database}, a rate-based automatic reaction network generator, has been shown to be applicable under both \ce{H2}-rich and \ce{H2O}-rich atmospheric conditions \citep{yang2024chemical}. Given that Jupiter’s equilibrium temperature falls within the range observed for temperate sub-Neptunes and considering its \ce{H2}-dominated atmosphere and potentially water-rich interior, we adopted the chemical reaction network from \cite{yang2024chemical} to perform one-dimensional thermochemical kinetic-transport modeling of Jupiter’s troposphere. This framework may also be suitable for modeling the atmospheres of Uranus and Neptune. The resulting chemical network includes 89 species involved in 1968 forward–reverse reaction pairs. It is provided as a YAML file (in \texttt{Cantera} format) and is available at: \url{https://doi.org/10.5281/zenodo.16750016} \citep{yang_2025_16750016}. Note that photochemical species and their associated reactions are not considered in the current study and have thus been excluded from the network. The comparison among various chemical networks used in previous 1D modeling studies of Jupiter's troposphere \citep{Wang_2016_Jupiter, cavalie2023subsolar} is shown in Figure~C\ref{fig:chemical_network_comparison} in Appendix \ref{sec:appendix_network_comaprisons}. In this study, for our one-dimensional thermochemical kinetic-transport atmospheric modeling (described in Section~\ref{subsec:epacris}), we investigated the effect of the Hidaka reaction on Jupiter’s deep atmospheric oxygen abundance by comparing both the conventional TST-based rate coefficient from \citet{moses2011disequilibrium} and the \textit{d}-TST-based coefficient from \citet{sanches2017novel}. For the two-dimensional hydrodynamic modeling (described in Section~\ref{subsec:snap}), we used only the rate coefficient reported by \citet{moses2011disequilibrium}.  

\subsection{1D Thermochemical Kinetic-transport Atmospheric Modeling$-$\rm \texttt{EPACRIS}} \label{subsec:epacris}
Based on the various elemental abundance scenarios described in Section \ref{subsec:elemental}, the $T$–$P$ and $K$\textsubscript{zz} profiles of Jupiter’s atmosphere described in Section \ref{fig:tp_kzz_profiles}, and the chemical network described in Section \ref{subsec:chemical_network}, we performed 1D thermochemical kinetic-transport modeling of Jupiter’s troposphere using the chemistry module of \texttt{EPACRIS} \citep{yang2024automated}. This model simulates the steady-state vertical mixing ratios of chemical species under different O/H scenarios in Jupiter’s atmosphere. As noted in Section \ref{subsec:inputs}, the primary goal of this study is to investigate the quenching of carbon monoxide in Jupiter’s deep atmosphere. Accordingly, photochemical reactions are not included in our study. This limits the validity of our results to pressures below at least 4 bars, above the \ce{H2O} cloud deck, where ultraviolet photons cannot penetrate\citep{bjoraker2018gas}. Since quenching occurs at much deeper levels, this exclusion of photochemistry does not affect our analysis.

\subsection{2D Hydrodynamic Modeling$-$\rm \texttt{SNAP}} \label{subsec:snap}
As the distribution of disequilibrium species is affected by cloud layer dynamics and deeper convective processes, we employ the \texttt{SNAP} hydrodynamical model used in \cite{hyder2025supersolar}, coupled with our newly constructed CO$-$\ce{CH4} chemical network generated by RMG (see Section~\ref{subsec:chemical_network} for details), to determine vertical distribution of CO in Jupiter's atmosphere. It employs a non-hydrostatic formulation with active cloud resolving capabilities \citep{SNAP2019}, which are crucial in simulating the distribution of trace species in convective and diffusive atmospheres.

The vertical coordinate of our hydrodynamic simulation spans the entire pressure range relevant to the problem, starting near $\sim 0.3$ and extending down to $\sim 7\times10^3$ bars (a vertical distance of $525$ km) with $\Delta z \sim 2$ km, encapsulating the water lifting condensation level (LCL) and the CO quench region in the deep atmosphere. The Jovian scale height, $H\sim40$ km, is well-resolved with this specification. The horizontal resolution is kept lower to limit the computational cost of the simulations with $\Delta x \sim 4$ km and spans $1200$ km, corresponding to about $1^{\circ}$ on Jupiter. Similar to \citet{hyder2025supersolar}, we allow for top cooling without bottom heating; the difference is only in the rate of numerical convergence and does not affect the results presented here. As the horizontal extent is small relative to the characteristic scale of Jupiter's zonal jets, large-scale latitudinal trends are ignored. This means that our results are primarily applicable to the equatorial region of Jupiter, although trace species have been observed to vary significantly in latitude \citep{Grassi+2020}. While our model is equatorially focused, the observational CO data used in this study include both equatorial and midlatitude measurements, consistent with Earth-based viewing geometries. Specifically, the original \citet{bezard2002carbon} measurement was taken from $6-12^{\circ}N$, while more recent observations from \citet{bjoraker2018gas} targeted the Great Red Spot, located near $20^{\circ}S$. The combined results from both studies were used to establish the observational range shown as the cross-hairs in Figure~\ref{fig:2D_hydrodynamic_modeling}. We apply a reflective and periodic boundary for the vertical and horizontal coordinates, respectively, with homogeneous cooling employed at the top boundary, identical to the setup in \cite{hyder2025supersolar}.

The simulations are evolved till the rate of change of the mean CO mole fraction, $\partial\langle X_{\rm CO}\rangle/\partial t$, in the top 20 bars of the atmosphere stabilizes to $\sim 0$. This occurs when this value falls below machine precision $(10^{-16})$ (Figure~E\ref{fig:f_prime_conv}), indicating a statistical steady state between the thermochemical production/loss terms and the vertical turbulent transport. We focus on the top 20 bars because this region encompasses the upper atmosphere where the mean vertical tracer gradient varies significantly, particularly near the water cloud deck, which lies between 5 and 10 bars depending on the O/H enrichment. The tracer gradient remains non-zero in this region, ensuring that any variability in tracer transport is fully captured in the average advection rate. The primary equations of motion that are solved in our model are the Euler equations, which are detailed by \citet{SNAP2019} along with their numerical treatment and not repeated here.

\subsection{Deriving empirical $\tau$\textsubscript{chem}$(T,P)$ for use in the \rm \texttt{SNAP}\it$-$2D hydrodynamic modeling} \label{subsec:t_chem_fitting}
Due to the stiffness and computational expense of the system, it is not yet feasible to fully couple chemical reaction kinetics with hydrodynamic modeling described in Section~\ref{subsec:snap}. Therefore, it is essential to derive the chemical timescale as a function of pressure ($P$) and temperature ($T$) for use in the \texttt{SNAP}-2D hydrodynamical modeling (Section~\ref{subsec:snap}). For consistency and easier comparison, we followed the procedure described by \cite{wang2015new} and fit the chemical timescale for carbon monoxide using the following fitting formula:
\begin{equation} 
\tau_{\mathrm{chem}}(T,P) \sim \alpha \cdot \exp\left(\frac{\beta}{T}\right) \cdot P^\gamma\,\; [\mathrm{s}]
\end{equation}
where $T$ and $P$ represent temperature [K] and pressure [bar], respectively, and $\alpha$, $\beta$, and $\gamma$ are fitted coefficients. It should be noted that this parameterization for estimating the chemical timescale of CO is similar to the approach used by \citet{zahnle2014methane}, enabling more accurate predictions of CO quenching across a wide range of conditions while still capturing the overall behavior of the full chemical network. 

The fitting was applied to 10 datasets spanning a broad range of O/H = 0.3$–$7$\times$ and $K$\textsubscript{zz} = 5$\times$10$^6$$-$10$^9$ [cm$^2$/s]. The conditions for each dataset are summarized in Table C\ref{tab:dataset_timescale} in Appendix \ref{sec:appendix_chemicaltimescale}. Figure C\ref{fig:timescalefitting} in Appendix \ref{sec:appendix_chemicaltimescale} shows the comparison between the 10 datasets and the fitted points, the residuals after fitting, and the comparison between the chemical timescale derived by \cite{wang2015new} (V12, redline) and the timescale fitted in this work using the RMG-generated network \citep{yang2024chemical} (green line). The fitted coefficients $\alpha$, $\beta$, and $\gamma$ for $\tau_{\mathrm{chem}}(T,P)$ plugged into the \texttt{SNAP}-2D hydrodynamic modeling in this study are:
\begin{align*} 
\alpha &= 2.89\times10^{-12} \\
\beta &= 41889.08 \\
\gamma &= -4.93\times10^{-7} \\
\end{align*}

% Recent hydrodynamical models that incorporate simplified disequilibrium thermochemistry have shown that dynamics near the cloud condensation level can substantially alter the vertical mean tracer distribution of the advected species.

\subsection{Chemical timescale approach approximation} \label{subsec:chemical_timescale_approach}
For comparison, we also show chemical timescale approaches using two distinct RMG-generated chemical networks: one for hot Jupiter atmospheres \citep{yang2024automated} and one for temperate sub-Neptune atmospheres \citep{yang2024chemical} (see Figure~D\ref{fig:timescalefitting}). Each chemical timescale was calculated using \texttt{Cantera}'s \texttt{Kinetics.net\_production\_rates} module \citep{Goodwin_Cantera_An_Object-oriented_2024} along the $T$–$P$ profile described in Section~\ref{subsec:inputs} to construct the Jacobian matrix of the chemical system. We then computed the eigenvectors and eigenvalues of the Jacobian matrix using \texttt{scipy.linalg.eig} \citep{2020SciPy-NMeth} and projected CO onto these eigenmodes. The mode with the largest projection was defined as the chemical timescale of CO, $\tau'_{\mathrm{chem}}(T,P)$, where the prime ($'$) distinguishes it from the empirical timescale derived in the previous Section~\ref{subsec:t_chem_fitting} (Recall that the empirical timescale is used in the 2D hydrodynamic modeling discussed in Sections~\ref{subsec:snap} and \ref{subsec:t_chem_fitting}). Further methodological details are provided in Appendix~\ref{sec:appendix_jacobian}. 

Previous chemical timescale approaches \citep[e.g.,][]{visscher2010deep, visscher2011quenching, moses2011disequilibrium, zahnle2014methane, Wang_2016_Jupiter} first incorporated all carbon reactions in 1D chemical-kinetic transport models, then isolated a single rate-limiting step to examine CO quenching behavior across different atmospheric regimes. While choosing a single reaction may simplify calculations, this assumption can lead to inaccuracies since multiple reactions often proceed at comparable rates, and slight variations in conditions (e.g., $T$, $P$, or chemical compositions) can alter which reaction is rate-limiting. Instead, the chemical timescale is derived directly from the eigenmodes and their corresponding eigenvalues of the Jacobian matrix at thermal equilibrium, explicitly considering all relevant reactions in which CO participates. This comprehensive approach provides more precise and realistic estimates for the quenched CO abundance.

The vertical mixing timescale is given by
\begin{equation}
\tau_{\mathrm{mix}} = \frac{L^2}{K_{\mathrm{zz}}},
\end{equation}
where $L$ is the characteristic mixing length, and $K_{\rm zz}=10^8$ [cm$^2$/s] is adopted as a nominal value. We adopt $L = 0.12H$ (where $H$ is the pressure and temperature-dependent scale height of Jupiter's atmosphere), based on mixing length theory \citep{smith1998estimation} and previous work on CO quenching kinetics in Jupiter-like atmospheres \citep{visscher2010deep}. It should be noted that this parameterization is primarily empirical rather than derived from first principles \citep{smith1998estimation}. We assume that CO quenches at the level where $\tau'_{\mathrm{chem}}(T,P) = \tau_{\mathrm{mix}}$, following the classical timescale comparison approach \citep{visscher2010deep}. Figure~\ref{fig:1D_vs_chemicaltimescale} compares the resulting quench levels for two end-member oxygen abundance scenarios (O/H = 2.3$\times$ and 0.3$\times Z_{\odot}$) using the chemical timescale approach to the outcomes from the \texttt{EPACRIS}$-$1D modeling results. A detailed discussion is provided in Section~\ref{subsubsec: 1D_vs_timescale_approach}.

\section{RESULTS AND DISCUSSIONS} \label{sec:results_discussion}

\subsection{Inclusion of the ``Hidaka reaction'' in the chemical network} \label{subsec:hidaka_reaction}
The Hidaka reaction refers to the thermal decomposition of methanol by a hydrogen radical:
\begin{equation}
    \ce{CH3OH}+\ce{H} \rightarrow \ce{CH3} + \ce{H2O}
\end{equation}
originally reported by the experimental study of \citet{hidaka1989thermal}. The reaction has been widely studied in the combustion society. The overall reaction \ce{CH3OH + H} proceeds via three primary channels:
(1) \ce{CH2OH + H2},
(2) \ce{CH3O + H2}, and
(3) \ce{CH3 + H2O},
with the third pathway (the Hidaka reaction) being the most exothermic, as confirmed by ab initio calculations \citep{lendvay1997ab, sanches2017novel}.

This Hidaka reaction has also been incorporated into chemical modeling of various (exo)planetary atmospheres \citep{visscher2010deep, venot2012chemical, wang2015new}. However, its inclusion has been a subject of debate. As noted by \citet{visscher2010deep} and \citet{moses2014chemical}, the rate coefficient reported by \citet{hidaka1989thermal} appears anomalously fast relative to its high energy barrier. In response, \citet{moses2011disequilibrium} recalculated the rate coefficient (referred to as nominal value) using high-level quantum chemical methods (QCISD(T)/CBS//QCISD(T)/cc-pVTZ) and transition state theory with torsional and asymmetric Eckart tunnelling corrections, while others chose to exclude the reaction from their chemical networks \citep{Wang_2016_Jupiter, venot2020new, cavalie2023subsolar}.

However, two important issues arise regarding how this reaction has been handled in prior works. First, the rate coefficient widely adopted by earlier studies \citep{visscher2010deep, venot2012chemical} was based on a misreported value in the NIST database \citep{NIST_2020}, rather than the original data from \citet{hidaka1989thermal}. Specifically, the NIST database (as of \today)\space lists the rate coefficient as 3.32$\times10^{-10}$ [cm$^3$/molecule/s] $e^{-22.20 [\rm kJ/mol]/RT}$, which is two orders of magnitude faster than the actual value reported by \citet{hidaka1989thermal}. The original expression by \citet{hidaka1989thermal} is 2.0$\times10^{12}$ [cm$^3$/mol/s] $e^{-5.3 [\rm kcal/mol]/RT}$. When converted to molecule-based units (i.e., by dividing by Avogadro’s number), the resulting expression becomes 3.32$\times10^{-12}$ [cm$^3$/molecule/s] $e^{-22.20 [\rm kJ/mol]/RT} $, which is consistent in activation energy (note that 1 cal = 4.184 J) but 100$\times$ slower in pre-exponential factor.

\begin{figure}[htb!]
    %\centering    
    \includegraphics[width=0.49\textwidth]{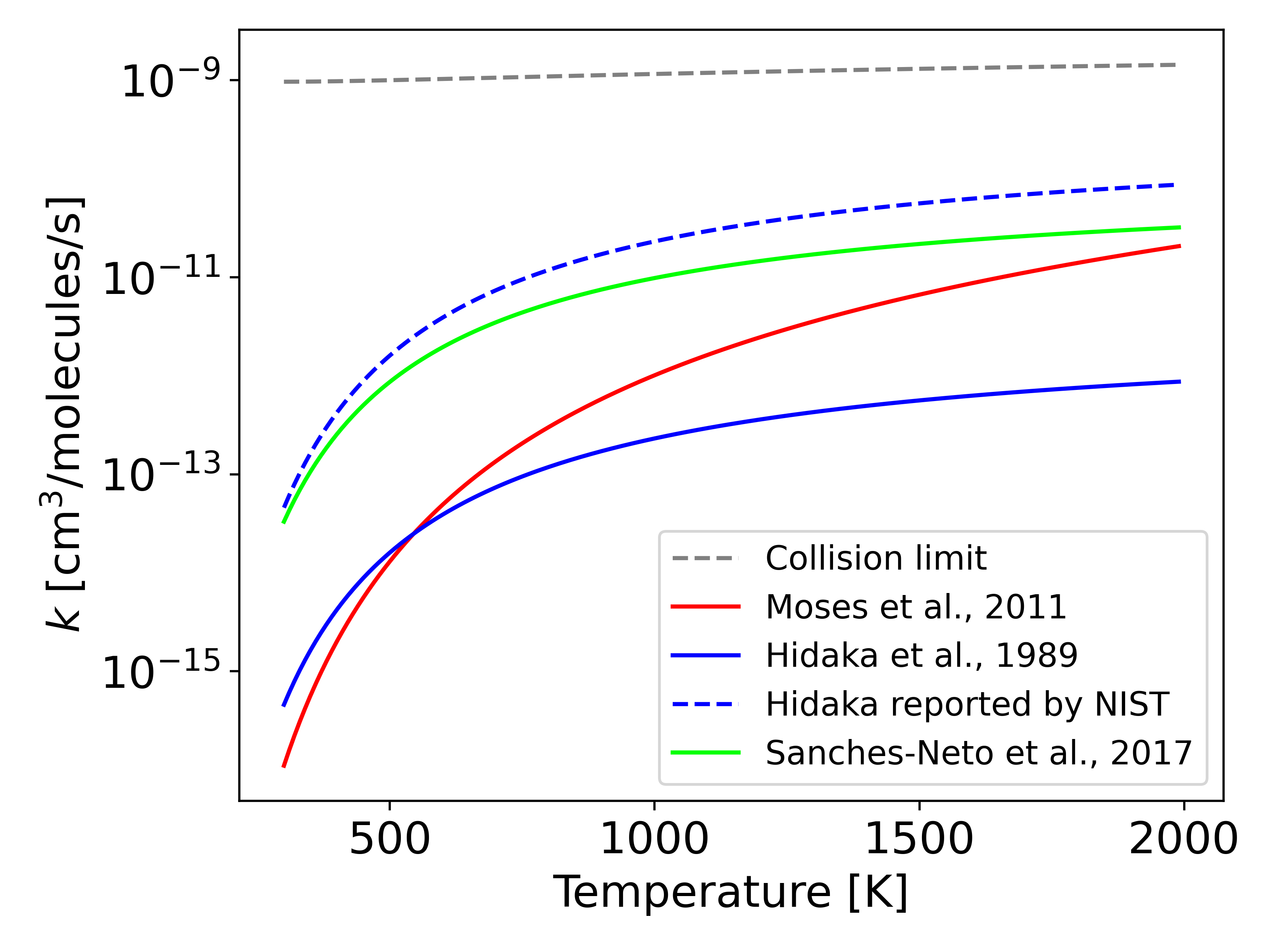}
     \caption{\footnotesize The temperature dependence of the rate coefficients for the reaction \ce{CH3OH + H} $\rightarrow$ \ce{CH3 + H2O} from various references. The solid blue line represents the original rate coefficient reported by \citet{hidaka1989thermal}; the blue dashed line corresponds to an incorrect version of this rate listed in the NIST database \citep{NIST_2020}; the red solid line shows the rate coefficient calculated by \citet{moses2011disequilibrium}; the lime solid line represents the rate calculated by \citet{sanches2017novel} using the \textit{d}-TST method; and the grey dotted line shows the collision limit calculated using the equation (1) from \citet{chen2017_collision_limit} for reference.}
    \label{fig:hidaka_reaction}
\end{figure}

As shown in Figure~\ref{fig:hidaka_reaction}, the corrected rate coefficient from \citet{hidaka1989thermal} is in reasonable agreement with the high-level quantum chemical rate calculated by \citet{moses2011disequilibrium} within an order of magnitude at temperatures below 1000 K. The discrepancy increases at higher temperatures (above 1000 K), where the Hidaka value becomes approximately more than an order of magnitude slower. This underscores the importance of verifying rate coefficient sources, particularly when extracting values from secondary databases such as the NIST database \citep{NIST_2020}. Inaccuracies in pre-exponential factors can have substantial effects on chemical network predictions.

Another issue regarding the treatment of the Hidaka reaction in prior studies is that some chemical networks have intentionally excluded this reaction from their chemical networks \citep{venot2020new, cavalie2023subsolar}. This exclusion was typically justified by the assumption that the competing hydrogen-abstraction channels forming \ce{CH2OH} + \ce{H2} and \ce{CH3O} + \ce{H2} dominate over the entire temperature range up to 2000 K, as supported by earlier studies \citep{hoyermann1981mechanism, li1996experimental, kerkeni2004ab, carvalho2008theoretical, meana2011high}. However, it is important to note that some of these prior works \citep[e.g.,][]{li1996experimental} did not consider the \ce{CH3} + \ce{H2O} channel (i.e., the Hidaka channel) in their analyses, and precise experimental measurement of branching ratios for the \ce{CH3OH} + H reaction remains challenging due to the complexity of methanol combustion, which involves numerous competing and side reactions.

Moreover, as shown in Fig. 5 of \citet{sanches2017novel}, experimental measurements of the total rate coefficient for the \ce{CH3OH} + H reaction reveal clear deviations from linear Arrhenius behavior at lower temperatures. This sub-Arrhenius curvature is commonly attributed to quantum tunneling effects, which become increasingly significant as temperature decreases \citep{silva2013uniform, carvalho2017deformed}. These features suggest the need for a more refined theoretical treatment in the moderate tunneling regime. To address this, \citet{sanches2017novel} performed an in-depth assessment of the Hidaka channel using ab-initio \textit{Car-Parrinello} molecular dynamics simulations at two representative temperatures (300 and 2500 K), followed by rate coefficient calculations using high-level \textit{deformed}-transition state theory (hereafter referred to as \textit{d}-TST). This approach offers improved treatment of quantum tunneling compared to conventional transition state theory, which had been used in previous calculations of the Hidaka channel's rate coefficient \citep{lendvay1997ab, moses2011disequilibrium}. \citet{sanches2017novel} demonstrated that \textit{d}-TST provides a more accurate description of the overall reaction rate constant for the \ce{CH3OH}+H reaction, particularly in the moderate tunneling regime (T$\leq$1000 K), compared to previous theoretical approaches \citep{lendvay1997ab, carvalho2008theoretical, moses2011disequilibrium, meana2011high}.

Surprisingly, as shown in Figure \ref{fig:hidaka_reaction}, the Hidaka reaction rate coefficient calculated using \textit{d}-TST by \citet{sanches2017novel} closely matches the misreported NIST value (lime solid vs. blue dashed line). At higher temperatures, the \textit{d}-TST rate aligns with the conventional TST rate \citep{moses2011disequilibrium} indicated by the red solid line, but becomes up to two orders of magnitude faster at temperatures near 300 K, reflecting significant quantum tunneling. Thus, using the \textit{d}-TST rate in one-dimensional kinetic-transport models lowers the predicted quenched CO mixing ratio due to enhanced CO-to-\ce{CH4} conversion. As shown in Figure B\ref{fig:chemical_network_comparison}, models excluding the Hidaka reaction (e.g., V20 or pre-2014 combustion networks) predict CO mixing ratios of $\sim4-5\times10^{-8}$ at 600 K. Including the conventional TST rate lowers this to $3-4\times10^{-8}$, and the \textit{d}-TST rate yields $\sim2\times10^{-8}$. 

As pointed out in \citet{Wang_2016_Jupiter}, the inclusion of the misreported NIST-database rate to V20 leads to a significantly lower value of $3-4\times10^{-9}$ (red solid line in Figure B\ref{fig:chemical_network_comparison}). This result highlights the high sensitivity of the manually built chemical network (V12) to this particular reaction, with the quenched CO abundance differing by approximately an order of magnitude between V12 and V20. In contrast, the RMG-generated network shows much less sensitivity: even with the fastest \textit{d}-TST Hidaka rate, the quenched CO value changes by less than a factor of two. This robustness highlights the advantage of RMG's rate-based and more comprehensive network construction over the manually constructed chemical network.

Given the significant role of the Hidaka reaction across the temperature range relevant to Jupiter’s deep atmosphere, and its validation by multiple \textit{ab initio} studies, this reaction, at least, should be included in chemical networks using either the \textit{d}-TST \citep{sanches2017novel} or conventional TST rate \citep{moses2011disequilibrium}. In this study, we adopt the conventional TST rate from \citet{moses2011disequilibrium} as the default. Omitting the Hidaka reaction introduces unnecessary uncertainty and can lead to underestimation of the quenched CO abundance, a key diagnostic for constraining Jupiter’s deep atmospheric oxygen content.

\subsection{Constraining the Oxygen Abundance using the 1D Thermochemical Modeling} \label{subsec:1D_modeling_results}
In this section, we present results from the \texttt{EPACRIS}$-$1D thermochemical kinetic-transport atmospheric modeling (as described in Section \ref{subsec:epacris}) and discuss how they can help us constrain the oxygen abundance in Jupiter’s deep atmosphere.

\begin{figure*}[htb!]
    %\centering    
    \includegraphics[width=1\textwidth]{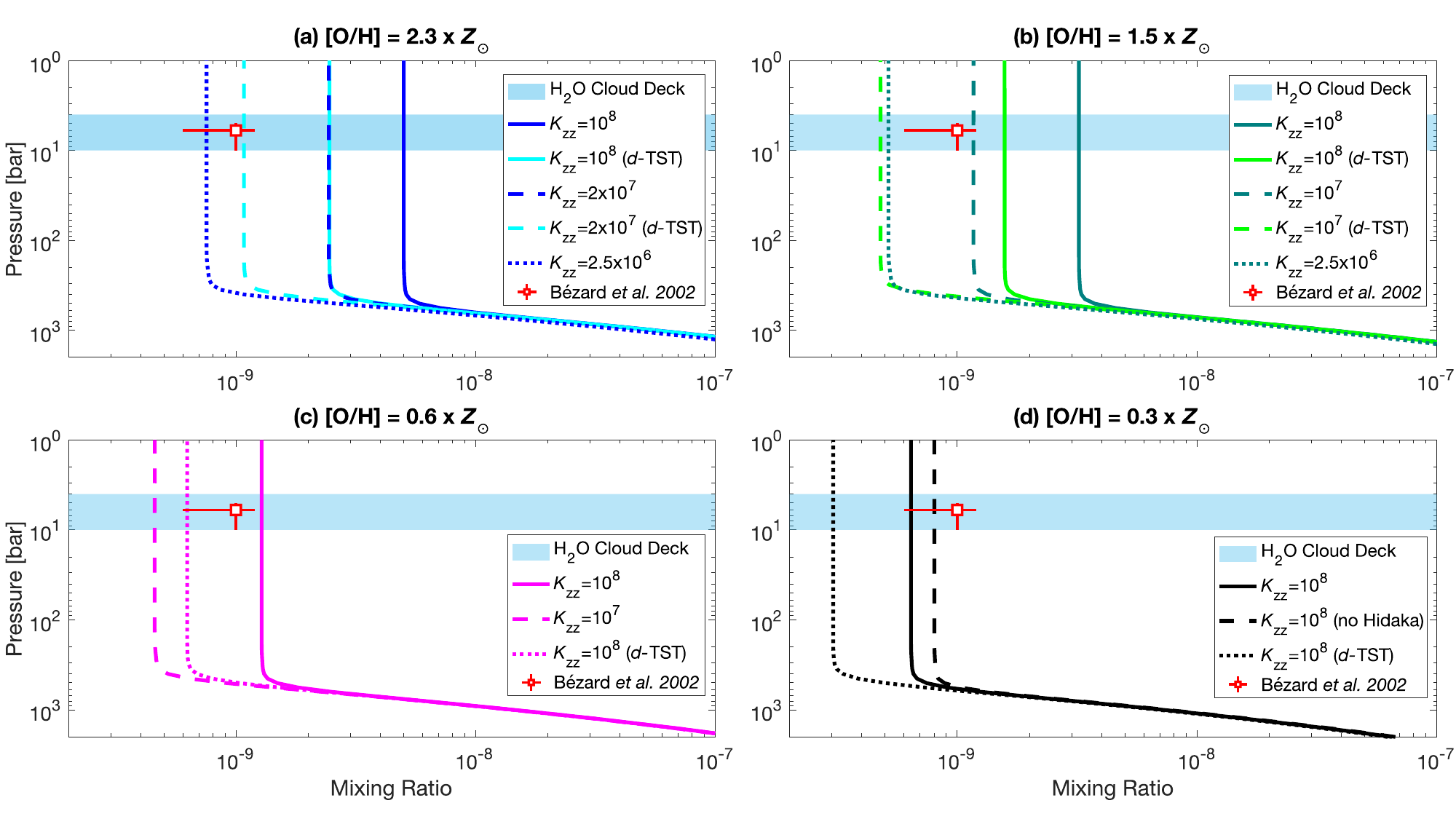}
     \caption{\footnotesize Vertical mixing‐ratio profiles of carbon monoxide ([CO]/[\ce{H2}]) in Jupiter’s atmosphere for various oxygen abundances O/H (Section~\ref{subsec:elemental}): 
(a) $2.3\times Z_{\odot}$, 
(b) $1.5\times Z_{\odot}$, 
(c) $0.6\times Z_{\odot}$, and 
(d) $0.3\times Z_{\odot}$. In each panel, we vary the eddy diffusion coefficient $K_{\rm zz}$ [cm$^2$/s] and adopt the Hidaka reaction rate coefficient from either \citet{moses2011disequilibrium} (nominal) or \citet{sanches2017novel} (indicated as \textit{d}-TST) Panel (d) additionally shows the CO profile simulated with the Hidaka reaction omitted from the chemical network described in Section \ref{subsec:chemical_network}. The red square with error bars indicates the observed upper‐tropospheric CO mixing ratio from \citet{bezard2002carbon}, with uncertainties from \citet{bjoraker2018gas}. The light blue shaded region indicates the water cloud decks between 4 and 10 bars.}
    \label{fig:1D_modeling}
\end{figure*}

\subsubsection{{\rm O/H} = $2.3\times Z_{\odot}$ scenario}
Among the various oxygen abundance scenarios, we first tested the 2.3$\times Z_{\odot}$ case close to the value of 2.7$\times Z_{\odot}$ inferred from Juno’s microwave radiometer data \citep{li2020water}. As shown in Figure~\ref{fig:1D_modeling}a, using an eddy diffusion coefficient of 10$^8$ cm$^2$/s, a commonly adopted value for Jupiter’s deep atmosphere \citep{visscher2010deep, wang2015new, Wang_2016_Jupiter, cavalie2023subsolar}, our model predicts a CO mixing ratio of ~5$\times$10$^{-9}$. This is almost a factor of five higher than the observed CO mixing ratio value reported by \citet{bezard2002carbon}. Even when using the \textit{d}-TST rate for the Hidaka reaction, which increases CO-to-\ce{CH4} conversion and lowers the CO abundance, the model still overpredicts CO by a factor of about two. To match the observed CO level, the eddy diffusion coefficient must be reduced to 2.5$\times$10$^6$ cm$^2$/s when using the nominal Hidaka rate \citep{moses2011disequilibrium}, or to 2$\times$10$^7$ cm$^2$/s when using the \textit{d}-TST rate \citep{sanches2017novel}. However, such low $K_{\rm zz}$ values, especially below $\sim10^6$ cm$^2$/s, are likely inconsistent with Jupiter’s observed thermal structure and intrinsic heat flux, which indicate vigorous convection in the deep troposphere, especially if one considers the dry Jupiter case in radiative-convective equilibrium \citep{Stone1976}, where the resulting eddy diffusion coefficient can be as high as $\sim10^9$ cm$^2$/s. However, a lower mixing strength is possible according to the 2D hydrodynamic modeling in this work (see Section~\ref{subsec:2D_modeling_results} and \ref{subsec:eddy_diffusion_coefficients}). These results suggest that an oxygen abundance of 2.3$\times Z_{\odot}$ is too high to reproduce the observed CO quenching behavior in Jupiter’s upper troposphere. Notably, Juno MWR’s latest analysis includes a lower bound of 1.5$\times Z_{\odot}$ \citep{LI2024116028}, allowing us to explore sub-2$\times Z_{\odot}$ scenarios.

\subsubsection{{\rm O/H} = $1.5\times Z_{\odot}$ scenario}
While the 1.5$\times Z_{\odot}$ oxygen abundance is introduced here in the context of the \texttt{EPACRIS}$-$1D thermochemical kinetic-transport modeling, this case was originally tested based on the 2D hydrodynamic modeling result using \texttt{SNAP}, which later identified it as best matching the observed CO mixing ratio \citep{bezard2002carbon} (see Section~\ref{subsec:2D_modeling_results}). As shown in Figure~\ref{fig:1D_modeling}b, similar to the 2.3$\times Z_{\odot}$ oxygen abundance scenario, using an eddy diffusion coefficient of 10$^8$ [cm$^2$/s] results in an overprediction of the CO mixing ratio: approximately a factor of two with the nominal Hidaka reaction rate and about a factor of three even when applying the \textit{d}-TST rate for Hidaka reaction. To match the observed CO level, the eddy diffusion coefficient needs to fall within approximately $3\times10^6$ to $1\times10^7$ cm$^2$/s when using the nominal Hidaka rate \citep{moses2011disequilibrium}, or between $1\times10^7$ and $\sim5\times10^7$ cm$^2$/s when using the \textit{d}-TST rate \citep{sanches2017novel}. This indicates that a lower $K_{zz}$ value is required in the deep atmosphere, consistent with the sluggish mixing implied by the presence of a stable radiative layer at depth, as suggested by \citet{cavalie2023subsolar}, to reproduce a supersolar oxygen abundance with their chemical network.

\subsubsection{{\rm O/H} = $0.6\times Z_{\odot}$ scenario}
An oxygen abundance of 0.6$\times Z_{\odot}$ provides the closest agreement with the observed CO mixing ratio \citep{bezard2002carbon} while maintaining the nominal eddy diffusion coefficient of 10$^8$ [cm$^2$/s]. As shown in Figure~\ref{fig:1D_modeling}, the CO mixing ratio predicted using the nominal Hidaka rate from \citet{moses2011disequilibrium} slightly exceeds the upper bound of the observational uncertainty \citep{bezard2002carbon}, whereas using the \textit{d}-TST rate yields a quenched CO mixing ratio near the lower bound. These results suggest that an oxygen abundance of 0.5$-$0.6$\times Z_{\odot}$ is consistent with the observed CO mixing ratio, though near the edges of the uncertainty range. 

Compared to the lower oxygen abundance of $\sim$0.3$\times Z_{\odot}$ reported by \citet{cavalie2023subsolar}, the present study favors a value roughly twice as high. This discrepancy likely stems from our inclusion of the Hidaka reaction and additional kinetic pathways absent from the V20 network used by \citet{cavalie2023subsolar}. Nevertheless, both the current 1D thermochemical modeling using \texttt{EPACRIS} and \citealt{cavalie2023subsolar} converge on a sub-solar oxygen abundance in Jupiter’s deep atmosphere, lending partial support to \citet{cavalie2023subsolar}’s conclusion regarding sub-solar oxygen levels in Jupiter’s troposphere, provided that a deep atmospheric eddy diffusion coefficient of 10$^8$ [cm$^2$/s] is assumed. However, this sub-solar value remains significantly lower than the lower bound of 1.5 $\times Z_{\odot}$ inferred from the latest Juno MWR analysis by \cite{LI2024116028}, suggesting that such low oxygen values are less likely to represent Jupiter’s true deep atmospheric composition.

\subsubsection{{\rm O/H} = $0.3\times Z_{\odot}$ scenario}
An oxygen abundance of $0.3\times Z_{\odot}$ was tested based on the previous 1D thermochemical kinetic-transport modeling by \citet{cavalie2023subsolar}, which identified around this value (i.e., $0.3\times$protosolar or $0.37\times Z_{\odot}$) as the best fit for reproducing the observed CO mixing ratio. As shown in Figure~\ref{fig:1D_modeling}d, adopting both the nominal eddy diffusion coefficient ($K_{\rm zz}$) of 10$^8$ [cm$^2$/s] and the nominal Hidaka rate coefficient \citep{moses2011disequilibrium} results in a CO mixing ratio near the lower bound of the observational uncertainty. In contrast, using the \textit{d}-TST rate leads to a predicted CO mixing ratio that falls significantly below the observed range. We also tested the effect of removing the Hidaka reaction from the chemical network, which shifted the quenched CO mixing ratio to $8\times10^{-10}$, a value that falls within the observed range, including its uncertainty, and is consistent with the result reported by \citet{cavalie2023subsolar}. 

These results suggest that a $K_{\rm zz}$ value larger than 10$^8$ [cm$^2$/s] is required to match the observed CO mixing ratio when assuming an oxygen abundance of 0.3$\times Z_{\odot}$. This higher $K_{\rm zz}$ value is less likely the case, since previous studies by \citet{wang2015new, Wang_2016_Jupiter} showed that $K_{\rm zz}$ decreases from about 10$^8$ [cm$^2$/s] near the equator to 10$^7$ [cm$^2$/s] at a latitude of 90$^{\circ}$. In addition, \citet{bjoraker2018gas} constrained an even lower \ce{GeH4} mixing ratio, which implies a required $K_{\rm zz}$ in the range of 5$\times10^6$ to 10$^8$ [cm$^2$/s], based on the \ce{GeH4} chemistry modeling by \citet{Wang_2016_Jupiter}. However, it should be noted that the chemical kinetics for \ce{GeH4} used in \citet{wang2015new} were based on analogies with silicon chemistry, and the actual behavior of germanium under Jovian conditions may differ. Therefore, improving our understanding of germanium kinetics remains an important direction for future study.

Overall, our \texttt{EPACRIS}$-$1D modeling across a range of O/H and $K_{\rm zz}$ values favors a best-fitting oxygen abundance between 0.5 and 0.6$\times Z_{\odot}$, with $K_{\rm zz}$ constrained to values below 10$^8$ [cm$^2$/s]. However, if slower mixing is allowed, as suggested by the presence of a deep and stable radiative layer in \citet{cavalie2023subsolar}, oxygen abundance as high as 1.5 to even 2.3$\times Z_{\odot}$ can match the observed CO mixing ratio \citep{bezard2002carbon}.

\begin{figure}[htb!]
    %\centering    
    \includegraphics[width=0.48\textwidth]{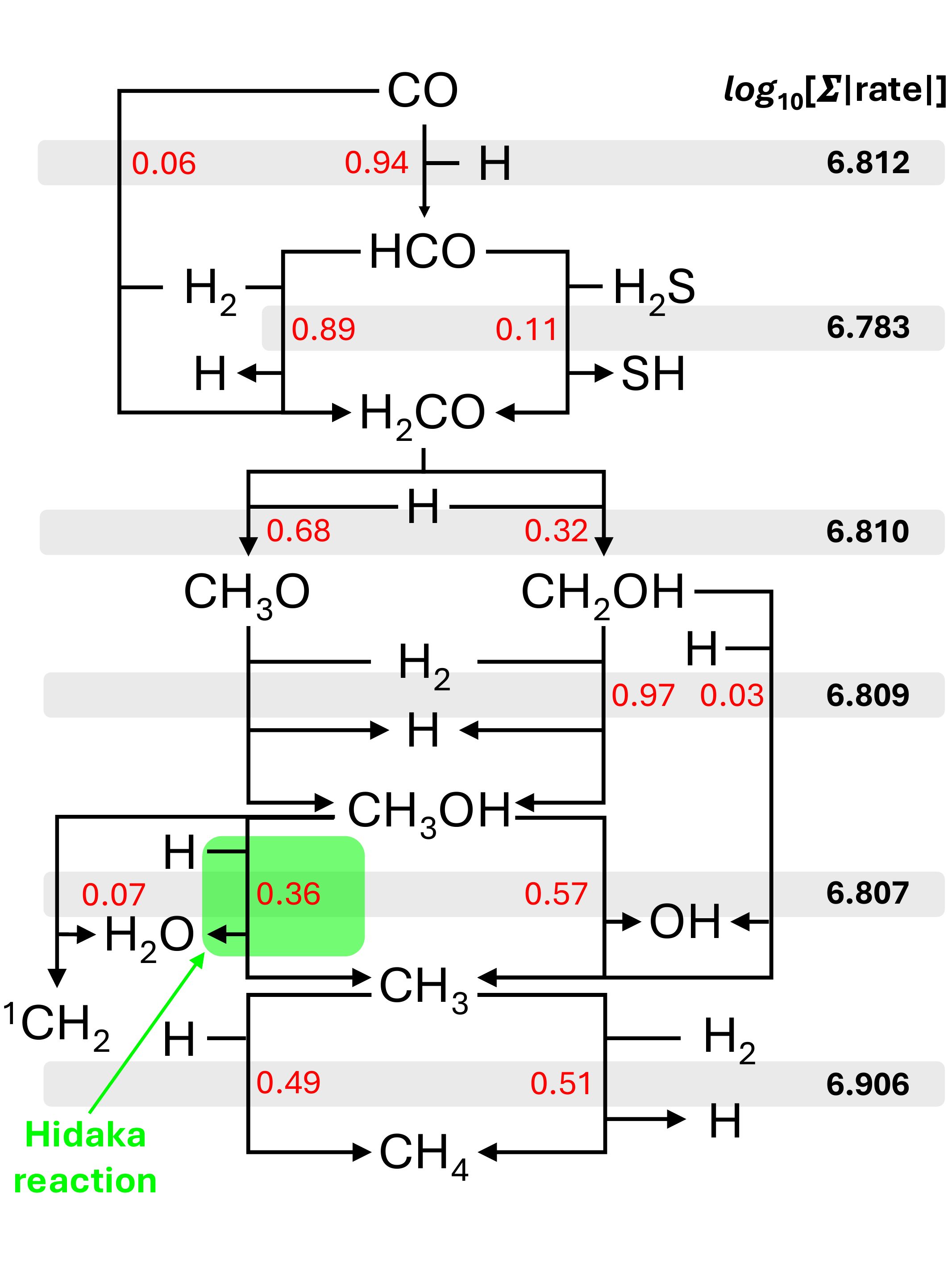}
     \caption{\footnotesize Major reaction pathways at the CO quenching point ($P \sim 505$ bar, $T \sim 1050$ K) for CO–\ce{CH4} conversion under O/H=$1.5\times Z_{\odot}$ and $K_{\rm zz}=10^{7}$ [cm$^2$/s]. Red numbers indicate branching ratios for species destruction. For example, 0.06 shows that 6\% of CO destruction occurs via the CO+\ce{H2}$\rightarrow$\ce{H2CO} reaction, with the remainder (94\%) occurring via CO+H$\rightarrow$HCO. Therefore, branching ratios sum to 1 (100\%). The black numbers on the far right indicate the logarithm of the absolute total rate [molecules/cm$^3$/s] of reaction pathways included in the light beige-shaded bars. For example, a value of 6.783 indicates a total HCO destruction rate of 10$^{6.783}$ [molecules/cm$^3$/s], representing the combined rates of the reactions HCO+\ce{H2}$\rightarrow$\ce{H2CO}+H and \ce{HCO}+\ce{H2S}$\rightarrow$\ce{H2CO}+SH. The green highlighted region identifies the Hidaka reaction, discussed in detail in Section \ref{subsec:hidaka_reaction}.}
    \label{fig:chemistry}
\end{figure}

\subsubsection{\ce{CO-CH4} chemistry in the deep atmosphere of Jupiter and chemical timescale approach}\label{subsubsec: 1D_vs_timescale_approach}
We performed a rate analysis of \ce{CO-CH4} chemistry to understand the \ce{CO-CH4} interconversion pathways in Jupiter's deep atmosphere in detail, as illustrated in Figure~\ref{fig:chemistry}. Although Figure~\ref{fig:chemistry} specifically depicts the reaction pathway scheme for CO quenching at O/H = 1.5$\times Z_{\odot}$ and $K_{\rm zz}=10^7$ cm$^2$/s, the primary chemical schemes were similar across three other oxygen abundance scenarios shown in Figure~\ref{fig:1D_modeling}, differing mainly in each species' total destruction rate.

Initially, 94\% of CO converts to \ce{HCO} via reaction with H radicals, with the remaining 6\% directly reacting with \ce{H2} to form \ce{H2CO}. Subsequently, \ce{HCO} predominantly reacts with \ce{H2} (89\%) or alternatively with \ce{H2S} (11\%) to yield \ce{H2CO}. \ce{H2CO} primarily interacts with H radicals, generating two isomers: \ce{CH3O} and \ce{CH2OH}. Both isomers mainly react with \ce{H2} to produce \ce{CH3OH}; however, \ce{CH2OH} also directly forms \ce{CH3} radicals through an additional reaction pathway involving H radicals. Then, 36\% of \ce{CH3OH} undergoes the Hidaka reaction, yielding \ce{CH3} and \ce{H2O}. The remainder decomposes via two unimolecular (thermal decomposition) pathways: forming \ce{CH3} + \ce{OH} (57\%) or \ce{^1CH2} + \ce{H2O} (7\%). Finally, \ce{CH3} reacts with either H radicals or \ce{H2} to produce \ce{CH4}, concluding the \ce{CO-CH4} interconversion process.

\begin{figure}[htb!]
    \renewcommand{\figurename}{Figure}
    %\centering    
    \includegraphics[width=0.48\textwidth]{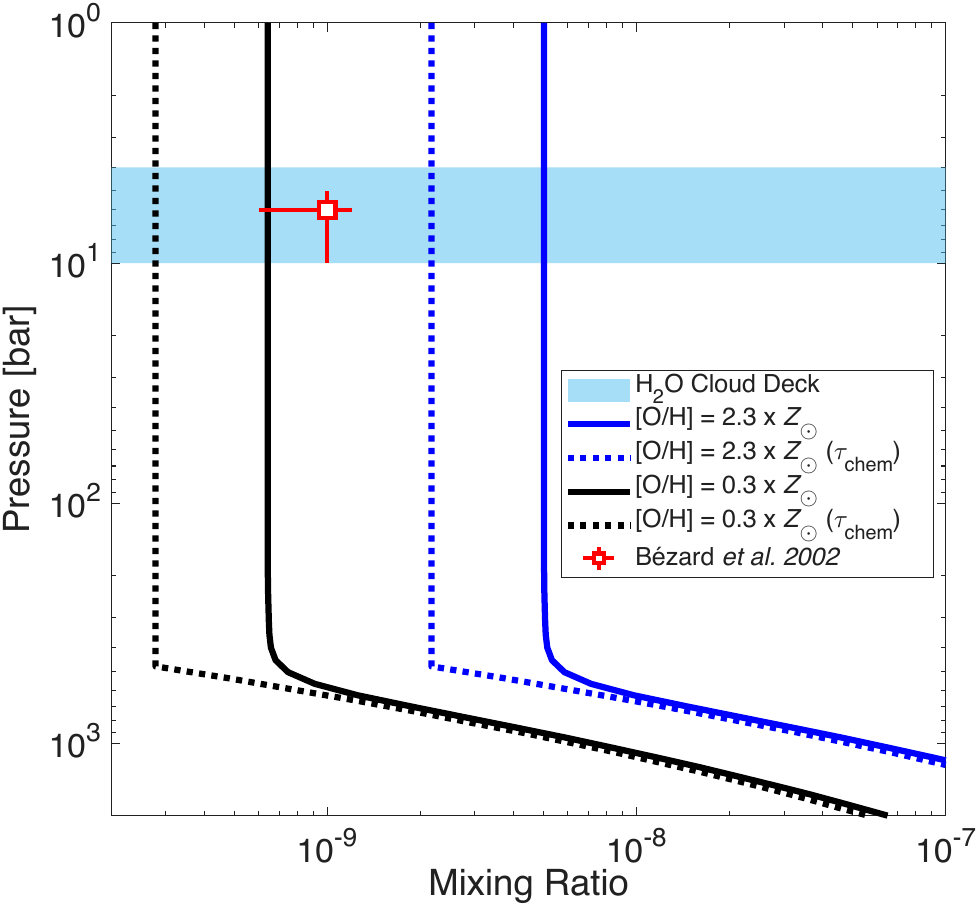}
     \caption{\footnotesize Comparison of \texttt{EPACRIS}$-$1D thermochemical simulations (solid lines; Section \ref{subsec:epacris}) with the chemical timescale approach approximation using the RMG-generated chemical network (dotted lines; method described in Section~\ref{subsec:chemical_timescale_approach}) for two oxygen abundance scenarios: O/H=2.3$\times Z_{\odot}$ (blue) and O/H=0.3$\times Z_{\odot}$ (black) assuming the same $K_{\rm zz}=10^{8}$ [cm$^2$/s]. The red square with error bars indicates the observed upper-tropospheric CO mixing ratio from \citet{bezard2002carbon}, with uncertainties from \citet{bjoraker2018gas}. The light blue shaded region indicates the water cloud decks between 4 and 10 bars.}
    \label{fig:1D_vs_chemicaltimescale}
\end{figure}

One notable point is the absence of a single representative rate-limiting step for the CO-to-\ce{CH4} conversion. Previous studies by \citet{visscher2010deep,visscher2011quenching} identified specific reactions, such as \ce{CH3OH}+H$\rightarrow$\ce{CH3O}+\ce{H2} or \ce{CH3OH}$\rightarrow$\ce{CH3}+OH, as rate-limiting step under Jupiter’s atmospheric conditions. Subsequently, \citet{wang2015new} used the rate coefficient for the latter reaction to constrain Jupiter's oxygen abundance within a range of $0.1–0.75\times Z_{\odot}$. However, our rate analysis (Figure~\ref{fig:chemistry}) demonstrates that no single reaction dominates. Instead, both the Hidaka reaction and the thermal decomposition of \ce{CH3OH} significantly contribute to \ce{CH3} formation.

Although we employed identical reaction rate coefficients to those used by \citet{visscher2011quenching} (Hidaka reaction from \citet{moses2011disequilibrium}; \ce{CH3OH} decomposition from \citet{jasper2007kinetics}), our analysis did not identify a single rate-limiting step, unlike their results. We argue that our finding is more realistic due to the complexity and highly coupled, nonlinear nature of the full chemical reaction network, which makes it unlikely for a single reaction to consistently dominate the rate across all conditions. Consequently, rather than approximating the chemical timescale by focusing on individual reactions \citep{visscher2010deep, visscher2011quenching, wang2015new}, it is more precise to directly compute the characteristic eigenmodes of the Jacobian matrix and their projection to the species of interest (e.g., CO in the current work) at thermal equilibrium across relevant $T$–$P$ profiles (e.g., the $T$-$P$ profile of Jupiter's atmosphere in the current work). This approach intrinsically captures the nonlinear interplay among all involved reactions and species. To accomplish this, we employed \texttt{Cantera} \citep{Goodwin_Cantera_An_Object-oriented_2024} (details described in Section~\ref{subsec:chemical_timescale_approach} and Appendix~\ref{sec:appendix_jacobian}).

While direct 1D chemical-kinetic transport modeling using \texttt{EPACRIS} is feasible for this study, its computational cost increases significantly with the size of the chemical network, as it scales linearly with the number of reactions and quadratically with the number of species \citep{schwer2002upgrading}. For large chemical networks, such as those developed for predicting polycyclic aromatic hydrocarbon formation, which involve 1594 species and 8924 reactions \citep{liu2021predicting}, the chemical timescale approach still remains a computationally practical alternative. This practicality is especially relevant for deep atmospheric studies of planets where thermochemical equilibrium chemistry dominates until it becomes comparable to vertical mixing, making it suitable for Jupiter, Uranus, and Neptune, as well as gas-rich exoplanets (e.g., sub-Neptunes and hot Jupiters). Our comparison of the chemical timescale with the full 1D modeling using \texttt{EPACIRS} (Figure~\ref{fig:1D_vs_chemicaltimescale}) showed that the chemical timescale approach underestimates the CO quenching mixing ratio by approximately 60\%. While discrepancies arise due to factors like assumed mixing length \citep[$L\sim0.12 H$;][]{smith1998estimation,visscher2010deep} and inherent method limitations (assuming quenching immediately happens at $\tau_{\rm chem}=\tau_{\rm mix}$), this deviation remains acceptable in exoplanetary atmospheric studies, where atmospheric retrieval uncertainties within an order of magnitude (1 dex) are typical and considered acceptable. Therefore, the chemical timescale approach may be especially suitable for future studies involving large chemical networks. 

\begin{figure*}[htbp]
  \centering
  % top‐row
  \begin{subfigure}[b]{0.495\textwidth}
    \centering
    \includegraphics[width=\textwidth]{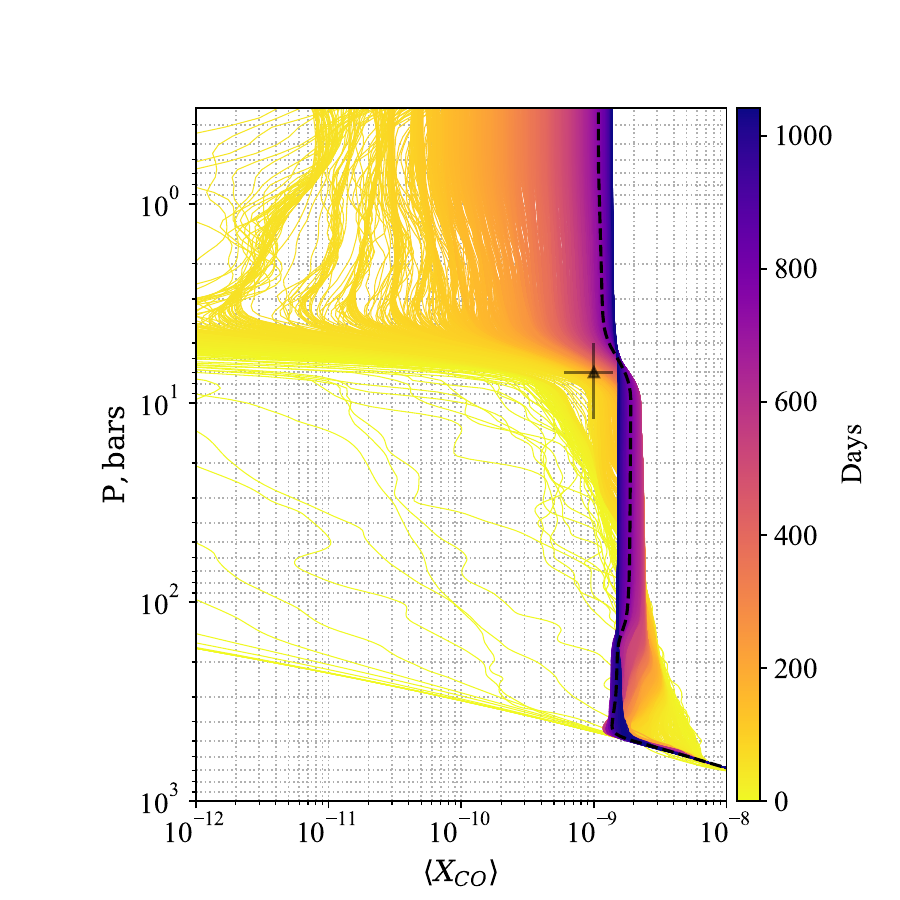}
    \caption{\textbf{O/H = 2.3$\times Z_{\odot}$}}
  \end{subfigure}
  \hfill
  \begin{subfigure}[b]{0.495\textwidth}
    \centering
    \includegraphics[width=\textwidth]{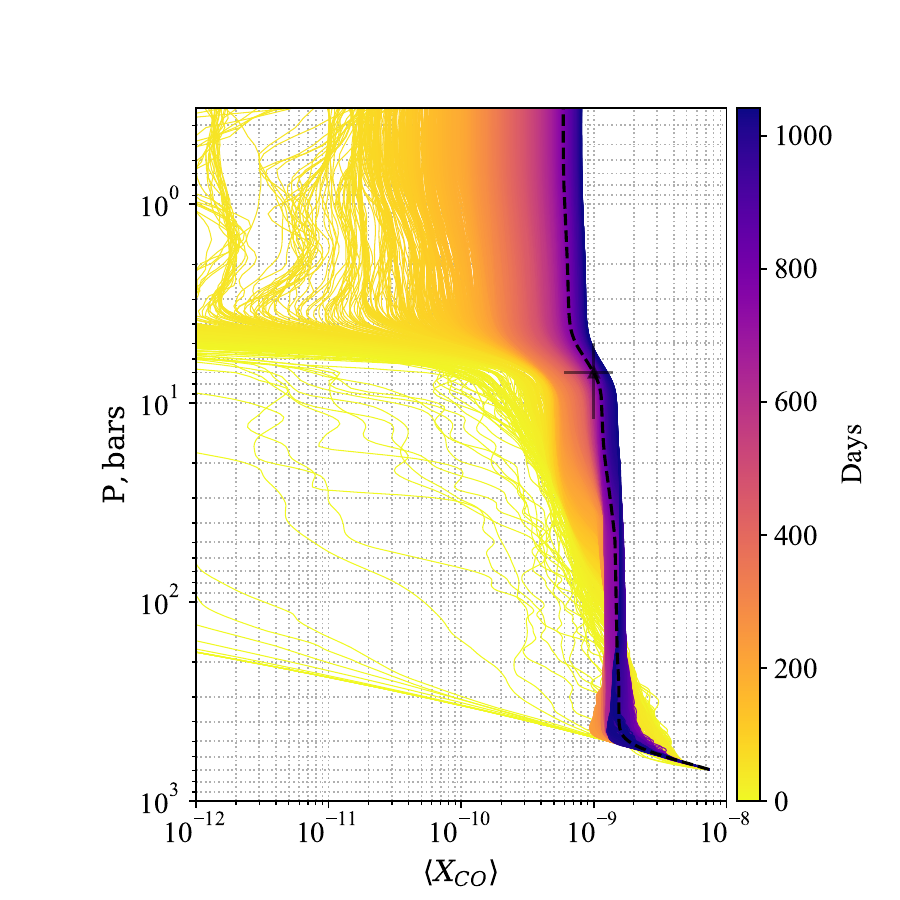}
    \caption{\textbf{O/H = 1.5$\times Z_{\odot}$}}
  \end{subfigure}

  \vspace{1em}  % adjust vertical gap

  % bottom‐row
  \begin{subfigure}[b]{0.495\textwidth}
    \centering
    \includegraphics[width=\textwidth]{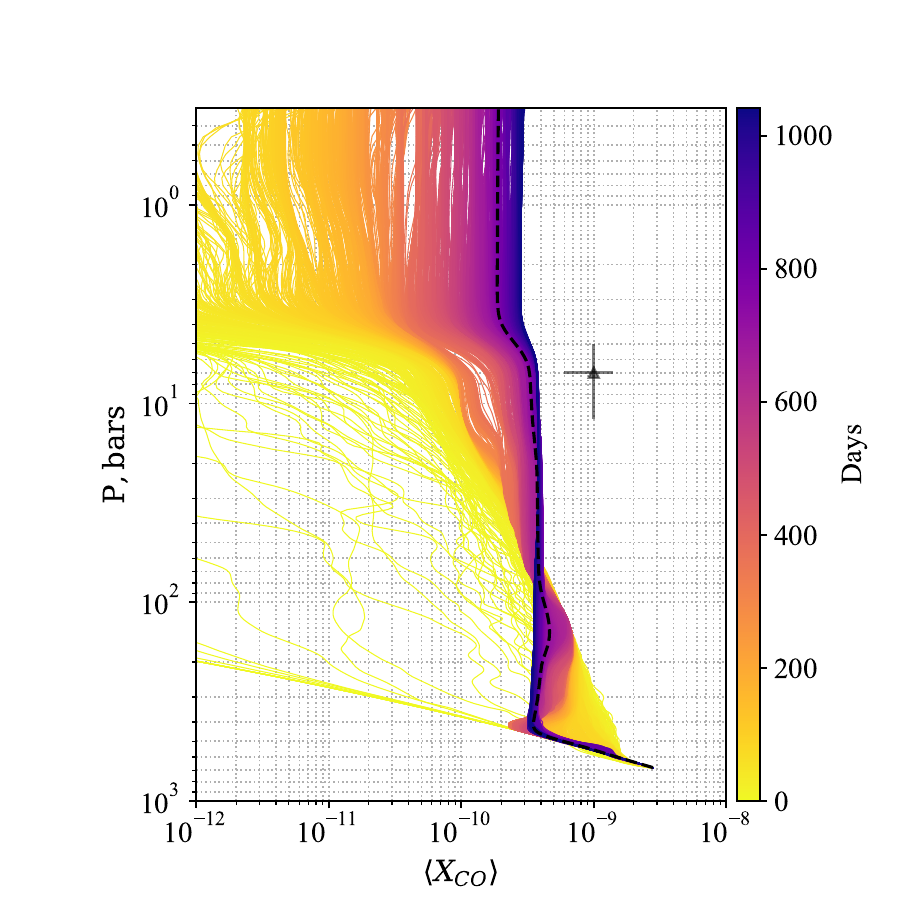}
    \caption{\textbf{O/H = 0.6$\times Z_{\odot}$}}
  \end{subfigure}
  \hfill
  \begin{subfigure}[b]{0.495\textwidth}
    \centering
    \includegraphics[width=\textwidth]{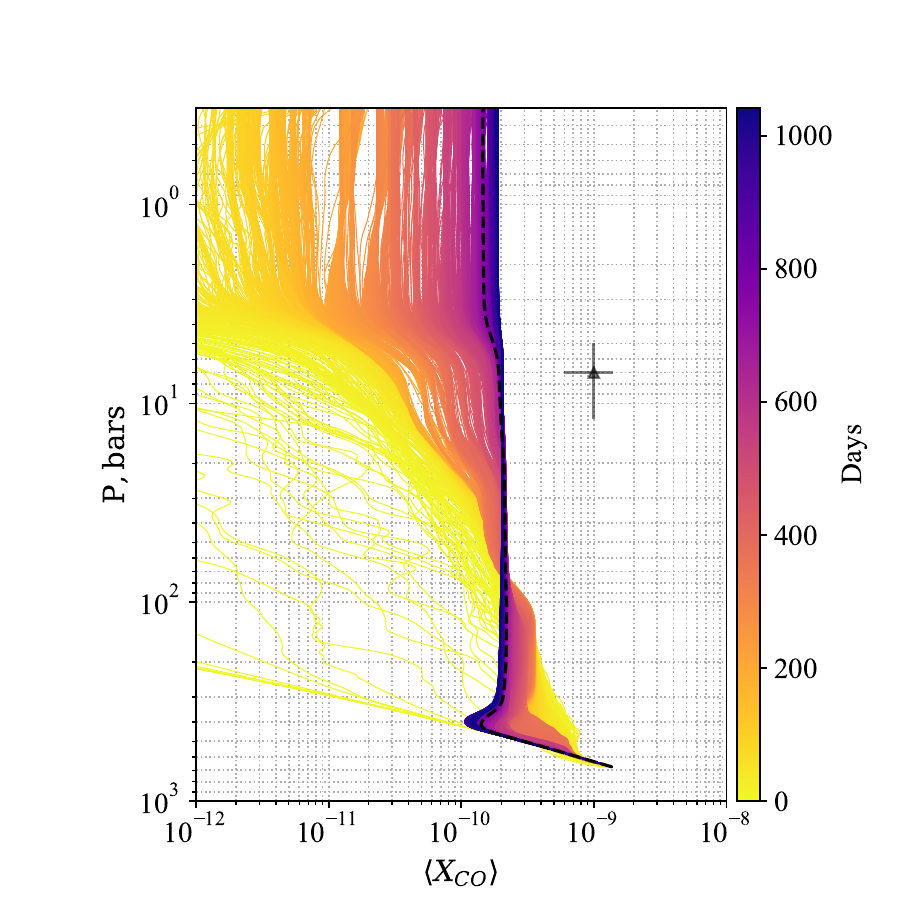}
    \caption{\textbf{O/H = 0.3$\times Z_{\odot}$}}
  \end{subfigure}

  \caption{The time-dependent behavior of the horizontally averaged CO mole fraction for four different oxygen abundance scenarios using \texttt{SNAP}$-$2D hydrodynamic modeling: (a) O/H = 2.3$\times Z_{\odot}$; (b) O/H = 1.5$\times Z_{\odot}$; (c) O/H = 0.6$\times Z_{\odot}$; and (d) O/H = 0.3$\times Z_{\odot}$. The solid gray triangle with error bars indicates the observed upper-tropospheric CO mole fraction from \citet{bezard2002carbon} and \citet{bjoraker2018gas}. Dashed lines represent the time-averaged CO mole fraction for each scenario. The underlying eddy diffusion strengths of these cases are not prescribed in the simulations and emerge naturally out of the dynamics. The eddy diffusion profiles are presented in Figure~\ref{fig:MED}.}
  \label{fig:2D_hydrodynamic_modeling}
\end{figure*}

\subsection{Constrainting the Oxygen Abundance using the 2D Hydrodynamic Modeling}\label{subsec:2D_modeling_results}
Compared to \citet{hyder2025supersolar}, the current simulations extend to nearly eight times longer run times, allowing for significantly improved convergence. This extended duration is critical because it ensures that the system evolves well beyond transient behavior and reaches a quasi-steady state. As a result, our outcomes provide a more robust and stable representation of the vertical distribution of atmospheric CO over long timescales as opposed to the short-term dynamics that investigated the transport of trace species across the LCL, which was a focus of \citet{hyder2025supersolar}. Running the previous \citet{hyder2025supersolar} simulation with a longer timescale would lower the inferred oxygen abundance from about 2.5$\times Z_{\odot}$ to a value closer to the 1.5$\times Z_{\odot}$. This value is still super-solar, so it remains consistent with our current results.

As shown in Figure~\ref{fig:2D_hydrodynamic_modeling}, once the \texttt{SNAP} simulations reach numerical convergence of 10$^{-16}$ (see Figure~E\ref{fig:f_prime_conv}), the vertical mixing ratio profiles of CO stabilize into nearly vertical trajectories across all oxygen abundance scenarios. As part of the convergence criteria, the vertical advection rate is observed to be below machine precision of 10$^{-16}$ (Section~\ref{subsec:snap} and Figure~E~\ref{fig:f_prime_conv}) for at least the latter half of the simulations. Among these scenarios, O/H = $1.5 \times Z_{\odot}$ best matches observational data \citep{bezard2002carbon, bjoraker2018gas} (see Figure~\ref{fig:2D_hydrodynamic_modeling}b). Additionally, Figure~E\ref{fig:1xsolar} shows results for O/H = $1 \times Z_{\odot}$, where the predicted CO mole fraction remains within observational uncertainty. Thus, our 2D hydrodynamic modeling using \texttt{SNAP} supports a modest supersolar oxygen abundance of $1$–$1.5 \times Z_{\odot}$, consistent with the lower bound of $1.5 \times Z_{\odot}$ derived from the latest Juno's MWR data analysis\citep{LI2024116028}. If the true oxygen abundance lies closer to the upper end of the Juno MWR data-inferred range (up to $8.3 \times Z_{\odot}$), which cannot be fit by our current model, this may indicate that additional mechanisms are acting to inhibit CO quenching deeper in the atmosphere. These may include processes beyond CO$–$\ce{CH4} chemistry and cloud microphysics, or possibly missing reactions within those frameworks. Clarifying this possibility will require more precise and extended observations of Jupiter's deep water reservoir.

As discussed in Sections~\ref{subsec:1D_modeling_results} and \ref{subsec:2D_modeling_results}, assuming O/H = $1.5 \times Z_{\odot}$ as inferred from our 2D hydrodynamic modeling using \texttt{SNAP}, the corresponding one-dimensional thermochemical kinetic-transport modeling yields an estimated eddy diffusion coefficient ($K_{\rm zz}$) in the range of $3\times 10^6$ to $5\times 10^7$ cm$^2$/s, depending on the choice of rate coefficient for the Hidaka reaction \citep{moses2011disequilibrium, sanches2017novel}. This value seems out of range compared to the commonly adopted value of $K_{\rm zz} = 10^8$ [cm$^2$/s], originally derived from \ce{GeH4} observations in Jupiter \citep{BJORAKER1986579}, and later constrained within $10^7$–$10^9$ [cm$^2$/s] by \citet{Wang_2016_Jupiter}, assuming factor-of-five uncertainties in the \ce{GeH4-GeO} chemical kinetics rates. However, a more recent analysis by \citet{bjoraker2018gas} reported an average \ce{GeH4} concentration of $0.35 \pm 0.05$ ppb in Jupiter’s Great Red Spot, approximately half the previously reported value ($0.7 \pm 0.2$ ppb). Using the analogous \ce{SiH4-SiO} chemistry approach from \citet{Wang_2016_Jupiter}, this updated measurement implies a revised $K_{\rm zz}$ in the range of $5 \times 10^6$–$10^8$ [cm$^2$/s], which now aligns well with our 1D model-derived $K_{\rm zz}$ range of $3 \times 10^6$–$5 \times 10^7$ [cm$^2$/s]) under the assumption of a $1.5 \times$ supersolar oxygen abundance.

It should be noted, however, that the analogy between \ce{GeH4-GeO} and \ce{SiH4-SiO} chemical kinetics is imperfect and requires further investigation. A more accurate constraint on $K_{\rm zz}$ would require updated kinetic data specific to the \ce{GeH4-GeO} system, which is beyond the scope of this study.

\subsection{Estimating $K_{\rm zz}$ from 2D hydrodynamic modeling using the Quasi Steady-State}\label{subsec:eddy_diffusion_coefficients}
Can we derive $K_{\rm zz}$ from 2D hydrodynamic modeling? Unlike the conventional 1D thermochemical kinetic-transport approach (here using \texttt{EPACRIS}), which often adopts a prescribed $K_{\rm zz}$ that is not firmly based on physical or chemical principles, 2D hydrodynamic modeling enables a more physics-based derivation. In this framework, the effective eddy diffusion coefficient, $K_{\rm zz}$, is not a free parameter but emerges from the underlying fluid dynamics. It depends on several factors, including the distortion of the chemical tracer field, making it a more self-consistent and physically motivated estimate of vertical mixing. Given the widespread use of $K_{\rm zz}$ in exoplanet atmospheric modeling, this approach offers a powerful framework that enables more robust constraints on $K_{\rm zz}$ based on physically grounded dynamics rather than arbitrary assumptions.

As our chemical equilibrium distribution is primarily a function of the vertical coordinate, it can be assumed to be horizontally uniform. Strictly speaking, this is dependent on the latitudinal distribution of water in the planet's atmosphere, whose non-uniformity would lead to additional non-diffusive effects \citep{Zhang&Showman2018}. In this capacity, our CO tracer field is horizontally well-mixed with a vertically varying interconversion time that becomes dominant near the 500 bar pressure region ($\tau_{\rm chem}\sim10^4$ s).

A general solution for the convection-diffusion-reaction equation is not known for arbitrary initial conditions. Historically, estimates of the mixing strength are derived assuming a dry atmosphere with Jupiter's thermal structure and heat flux \citep[e.g.,][]{Stone1976, FLASAR1986280}. Although these provide a useful constraint on the atmospheric turbulent diffusivity, they do not take into consideration the non-linear advective motions in the atmosphere and inhibitive effects near the cloud layer, which can cause a notable drop in the diffusive strength due to the decreased mass flux across the clouds \citep{Sugiyama+2014}. Although $K_{\rm zz}$ approximates the dominant vertical diffusive component in the context of dry convection, eddy diffusion is generally a tensor with non-analytical solutions for its terms.

\begin{figure}[htb!]
    \renewcommand{\figurename}{Figure}
    \centering    
    \includegraphics[width=0.49\textwidth]{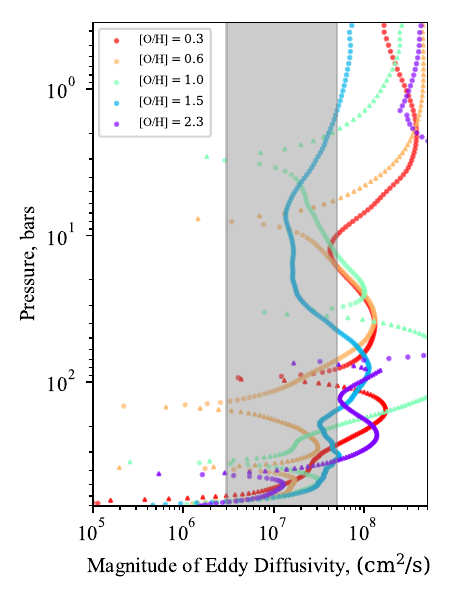}
     \caption{\footnotesize The magnitude of the mean eddy diffusion profiles as estimated using equation \ref{eq:kzz_full_ss}. The profiles are not positive-definite, as is the case within the 1D framework (\texttt{EPACRIS}). The positive values are demarcated by circles, while triangles are used for the negative values. The gray region represents the $3\times10^6$ to $5\times10^7$ [cm$^2$/s] range, which aligns with the $K_{\rm zz}$ values estimated from the \texttt{EPACRIS}$-$1D modeling under the assumption of $\rm O/H=1.5$ to match the observed CO abundance. The negative diffusivity is representative of the non-diffusive effects present in all of our simulations. The profiles occasionally become very large in magnitude as the tracer gradients get closer to zero, signifying a general breakdown of the fully diffusive treatment. The overall magnitudes, however, tend to be consistently within the range of expected vertical eddy coefficients that are estimated from 1D thermochemical modeling using \texttt{EPACRIS}.}
    \label{fig:MED}
\end{figure}

Although the full convection$-$diffusion$-$reaction equation has no general solution for arbitrary initial and boundary conditions \citep{Kim2020}, we derive an eddy diffusion coefficient from the steady-state conditions of our 2D hydrodynamic simulations using \texttt{SNAP}. Using the convection$-$diffusion equation from \citet{Vallis2017}, the eddy diffusion coefficient can be written as
\begin{equation}\label{eq:kzz_full_ss}
\overline{K_{\rm zz}} \sim \left( -\overline{v(z)q(z)} + \int_{z}{\overline{S(z')}dz'} \right)\left(\frac{\partial \overline{q}}{\partial z}\right)^{-1},
\end{equation}

\noindent where $v(z)$ is the vertical velocity, $q(z)$ is the CO vertical distribution, and $S(z')$ is the thermochemical term that approximates the CO production and loss as a function of the pressure and temperature distribution (the integration is performed over a dummy variable, $z'$). The overline indicates a horizontal and temporal average of the specific field. Since the eddy fluxes dominate in the vertical direction, $\overline{vq}\sim \overline{v'q'}$, resulting in the conventional relation between the mean tracer gradient distribution and the vertical eddy flux if the source term is ignored. Equation \ref{eq:kzz_full_ss} assumes a steady-state of the system, where $q(z)$ has stabilized and $\partial q(z)/\partial z$ is very close to 0 through most of the atmosphere. If one uses the actual steady-state tracer distribution (demarcated by the black dashed lines in Figure \ref{fig:2D_hydrodynamic_modeling}), equation \ref{eq:kzz_full_ss} becomes undefined in specific regions; specifically, the diffusive process is less dominant by the time the tracer field has fully stabilized. We use a mean tracer profile from the final few stages of the simulations that are approaching the steady-state distribution but still allow for the \emph{quasi} steady-state mean eddy diffusivity ($\overline{K_{\rm zz}}$) to be approximated. We present the corresponding $K_{\rm zz}$–$P$ profiles for each case shown in Figure~\ref{fig:2D_hydrodynamic_modeling} in Figure~\ref{fig:MED}.

We note that the eddy diffusivity occasionally becomes negative, which is consistent with the 3D hydrodynamical simulations performed by \citet{Zhang&Showman2018}. Depending on the correlation between the vertical velocity field and horizontal tracer fluctuation, the tracer field can be distorted large enough that the derived $\overline{K_{\rm zz}}$ falls below zero. This occurs for all oxygen enrichment cases as there is a significant non-diffusive contribution that shapes the tracer distribution. The mean horizontal eddy fluxes in our simulations are about 2 orders of magnitude larger than the vertical fluxes.

The interpretation of the diffusivity estimate is made more complicated by the fact that we use a varying CO timescale in contrast to the simplified short- and long-lived interconversion timescales employed by previous GCMs \citep{Zhang&Showman2018}. Thus, our simulations always retain some degree of non-diffusive contributions that shape the tracer distribution, particularly as the mean zonal flow can drastically alter the overall eddy diffusivity along the isobars. The magnitude of the vertical eddy diffusivity in each enrichment case tends to be around the expected range. Although these magnitudes exceed realistic bounds when a purely diffusive treatment is applied (as in equation \ref{eq:kzz_full_ss}) to regions with non-diffusive dynamics, the overall magnitudes never fall below $\sim10^5$ [cm$^2$/s] except for the $[\rm O/H]=0.3$ case, which falls as low as $\sim2\times10^4$ [cm$^2$/s] due to weakened mixing that is expected in such low humidity systems \citep{Seeley&Wordsworth2023}. Further exploration of the tracer distribution as a function of the vertical and horizontal eddy fluxes will be presented in a 3D model with latitudinal variability \citep{Hyder_3D_model}.

However, in general, as shown in Figure~\ref{fig:MED}, this 2D-hydrodynamic modeling-derived $K_{\rm zz}-P$ aligns well with our \texttt{EPACIRS}-1D modeling constraint on $K_{\rm zz}-P$ ($3 \times 10^6$–$5 \times 10^7$ [cm$^2$/s]) assuming a $1.5 \times$ supersolar oxygen abundance). This agreement is notable, given that the two modeling frameworks address complementary aspects of the problem. The \texttt{EPACRIS}$-$1D model captures detailed chemical kinetics but does not fully represent fluid dynamical processes. In contrast, the 2D hydrodynamic model (\texttt{SNAP}) resolves atmospheric motions but relies on simplified chemistry. By coupling these two approaches, our current framework offers a more comprehensive and first-principle grounded method for constraining both chemical and dynamical behavior in (exo)planetary atmospheres.

\subsection{Discussion of Elevated Jovian C/O Ratio in the Context of Planet Formation and Solar Nebula Chemistry}
The current work suggests an O/H=1-1.5$\times$solar abundance (corresponding to an \ce{H2O} abundance of $\sim$930--1395 ppm), which, when coupled with the firmly established GPMS measurement of Jupiter's [C/H]=4.0$\times$solar abundance, implies a significantly elevated planetary C/O$\sim$2.9 even compared to the recently proposed protosolar ratio ($0.73\pm0.10$) by \citet{Truong_2024}, which is higher than the commonly adopted value of 0.54 \citep{lodders2021relative}.  \citet{Truong_2024} argue that higher solar C/O ratios help reconcile the compositions of large Kuiper Belt Objects (KBOs), such as the Pluto–Charon system, by reducing water ice abundance through increased partitioning of carbon into refractory organics, CO, and \ce{CO2}. Our results might extend this narrative, demonstrating that Jupiter's formation region was likely far richer in carbon relative to oxygen compared to the broader protosolar nebula, highlighting significant local compositional heterogeneity and radial fractionation within the protoplanetary disk.
Such an elevated planetary C/O ratio suggests Jupiter predominantly accreted carbon-rich planetesimals or ices rather than water-rich solids, consistent with the reduced water ice predictions from the higher protosolar C/O scenario suggested by \citet{lodders2004jupiter, Truong_2024}. This modeling result reinforces a formation scenario where radial drift \citep{lunine2004origin, booth2017chemical}, differential condensation \citep{oberg2016excess}, or selective migration of carbon-rich solids substantially enriched Jupiter’s feeding zone. Together, these insights highlight the complexity and chemical diversity of the early solar nebula, providing essential benchmarks for interpreting compositional diversity within our solar system (which can be extended to Uranus and Neptune) and across exoplanetary systems. Further studies are required to clarify the mechanisms behind the preferential accretion of carbon-rich material, which will enhance our understanding of planetary formation and nebular chemistry.

\section{Conclusions} \label{sec:conclusions}

In this work, we have explored Jupiter's deep atmospheric oxygen abundance by coupling one-dimensional thermochemical kinetic-transport modeling and two-dimensional hydrodynamic modeling. Using a comprehensive and rate-based chemical network generated by the Reaction Mechanism Generator (RMG), we investigated the impact of key chemical pathways, particularly reviewing in detail and emphasizing the significance of the Hidaka reaction, which previous studies have inconsistently or improperly considered.

Our 1D modeling using \texttt{EPACRIS} suggests that an oxygen abundance ranging from approximately 0.5 to 0.6 times solar is most consistent with observational constraints, assuming standard eddy diffusion values ($K_{\rm zz}\leq10^8$ [cm$^2$/s]). However, such sub-solar oxygen abundances are less likely to represent Jupiter’s true deep atmospheric composition when compared to the latest Juno MWR analysis \citep{LI2024116028}. This discrepancy might suggest that slower vertical mixing is present, possibly due to stable radiative layers as proposed by \citet{cavalie2023subsolar}. Under such conditions, oxygen abundances as high as 2.3 times solar, or even higher, could also be consistent with the observational data.

In contrast, our 2D hydrodynamic simulations using \texttt{SNAP} strongly support a modest supersolar oxygen abundance of approximately 1$-$1.5 times solar, consistent with the lower bound inferred from the latest Juno MWR analysis \citep{LI2024116028}. We also introduce a method for deriving Jupiter's eddy diffusion coefficient ($K_{\rm zz}$) from 2D hydrodynamic modeling using the quasi-steady-state approach. Interestingly, rather than the nominal value of $K_{\mathrm{zz}} = 1\times10^8$ cm$^2$ s$^{-1}$, our results support a revised value of $K_{\mathrm{zz}} \sim 3\times10^6$ cm$^2$ s$^{-1}$, based on the agreement between two independent yet sophisticated approaches—1D chemical–kinetic transport modeling and 2D hydrodynamic modeling. This method has broad applicability to exoplanet atmospheric modeling, where $K_{\rm zz}$ remains highly uncertain despite its significant influence on atmospheric chemistry.

Our results also imply a notably elevated carbon-to-oxygen (C/O) ratio in Jupiter compared to current protosolar estimates (up to 0.73), suggesting that Jupiter likely accreted predominantly carbon-rich planetesimals or ices \citep{lodders2004jupiter}. This finding underscores the need for further study to clarify the mechanisms behind the preferential accretion of carbon-rich material during planetary formation, which will significantly inform models of giant planet formation and nebular chemistry.

Ultimately, our integrated modeling approach, which combines detailed thermochemical kinetics with hydrodynamic processes, provides a robust framework for constraining Jupiter’s deep atmospheric composition. This methodology not only advances our understanding of Jupiter but also serves as a comprehensive and physically grounded tool for investigating coupled chemical and dynamical processes in (exo)planetary atmospheres.

%% Please use the acknowledgment and contribution environments. This will 
%% be anonomyized when the "anonymous" style option is used. 
\begin{acknowledgments}
 The research was carried out at the Jet Propulsion Laboratory, California Institute of Technology, under a contract with the National Aeronautics and Space Administration (80NM0018D0004). JY was funded by the Caltech-JPL President's and Director's Research and Development Fund, awarded to RH. AH was funded by the NASA Postdoctoral Program (NPP). © 2025. California Institute of Technology. Government sponsorship. acknowledged. The High Performance Computing resources used in this investigation were provided by funding from the JPL Information and Technology Solutions Directorate.
\end{acknowledgments}
\begin{contribution}
JY conceptualized and led the project. JY, AH, and RH jointly designed the study. JY led the writing of the manuscript. JY conducted and analyzed 1D thermochemical kinetic-transport modeling (\texttt{EPACRIS}). AH conducted and analyzed 2D hydrodynamic modeling (\texttt{SNAP}). All authors contributed to the writing and revision of the manuscript.
\end{contribution}

\appendix
\setcounter{figure}{0}
\setcounter{table}{0}

\section{Elemental Parameterization of Jupiter's Atmosphere}  \label{sec:appendix_elements}
This appendix demonstrates a detailed description of the O/H variation mentioned in Section~\ref{subsec:elemental} in the form of Table A~\ref{tab:elemental_parameterization} for easier comparison with previous studies.

\begin{table*}[ht]
\renewcommand{\tablename}{Table A}
\centering
\caption{Elemental composition profiles used in this study to model Jupiter's atmosphere}\label{tab:elemental_parameterization}

\begin{tabular*}{\textwidth}{@{\extracolsep\fill}cccccc}
\toprule
& \multicolumn{5}{c}{\textbf{O/H enrichment compared to present-day solar abundance ($Z_\odot$)}}\\
$E$ & 0.3$\times$ & 0.6$\times$ & 1.0$\times$ & 1.5$\times$ & 2.3$\times$ \\
\midrule
\textbf{H}  & 0.925735 & 0.925597 & 0.925413 & 0.925183 & 0.924806 \\
\textbf{He (0.9$\times Z_\odot$)\textsuperscript{\textit{a}}} & 0.072670 & 0.072659 & 0.072645 & 0.072627 & 0.072597 \\
\textbf{C (4.0$\times Z_\odot$)\textsuperscript{\textit{b}}}  & 0.001097 & 0.001097 & 0.001097 & 0.001096 & 0.001096 \\
\textbf{N (4.7$\times Z_\odot$\textsuperscript{\textit{c}})}  & 0.000307 & 0.000307 & 0.000307 & 0.000307 & 0.000307 \\
\textbf{O\textsuperscript{\textit{d}}}  & 0.000149 & 0.000298 & 0.000497 & 0.000745 & 0.001153 \\
\textbf{S (3.2$\times Z_\odot$)\textsuperscript{\textit{e}}}  & 0.000041 & 0.000041 & 0.000041 & 0.000041 & 0.000041 \\
\textbf{C/O}& 7.36 & 3.68 & 2.21 & 1.47 & 0.95 \\
\botrule
\end{tabular*}
\tablecomments{\footnotesize\textsuperscript{\textit{a}} [He]/[\ce{H2}]=0.157 was taken from \cite{Wang_2016_Jupiter}. Thus He/H$\sim7.85\times10^{-2}$\\ \textsuperscript{\textit{b}} [\ce{CH4}]/[\ce{H2}]=2.37$\times10^{-3}$ was taken from \cite{Wang_2016_Jupiter}. Thus C/H$\sim1.19\times10^{-3}$ \\ \textsuperscript{\textit{c}} [\ce{NH3}]/[\ce{H2}]=6.64$\times10^{-4}$ was taken from \cite{Wang_2016_Jupiter}. Thus N/H$\sim3.32\times10^{-4}$\\ \textsuperscript{\textit{d}} O/H was varied by $A(E)$=8.73+log$_{10}E$, where $E$ is an enrichment factor for oxygen. Thus O/H$\sim10^{A(E)-12}$.\\ \textsuperscript{\textit{e}} [\ce{H2S}]/[\ce{H2}]=8.90$\times10^{-5}$ was taken from \cite{Wang_2016_Jupiter}. Thus S/H$\sim4.45\times10^{-5}$}
\end{table*}

%\FloatBarrier

\section{$T-P$ \& $K_{\rm zz}$ profiles}  \label{sec:appendix_inputs}
\begin{figure}[htb!]
    \renewcommand{\figurename}{Figure B}
    %\centering    
    \includegraphics[width=0.49\textwidth]{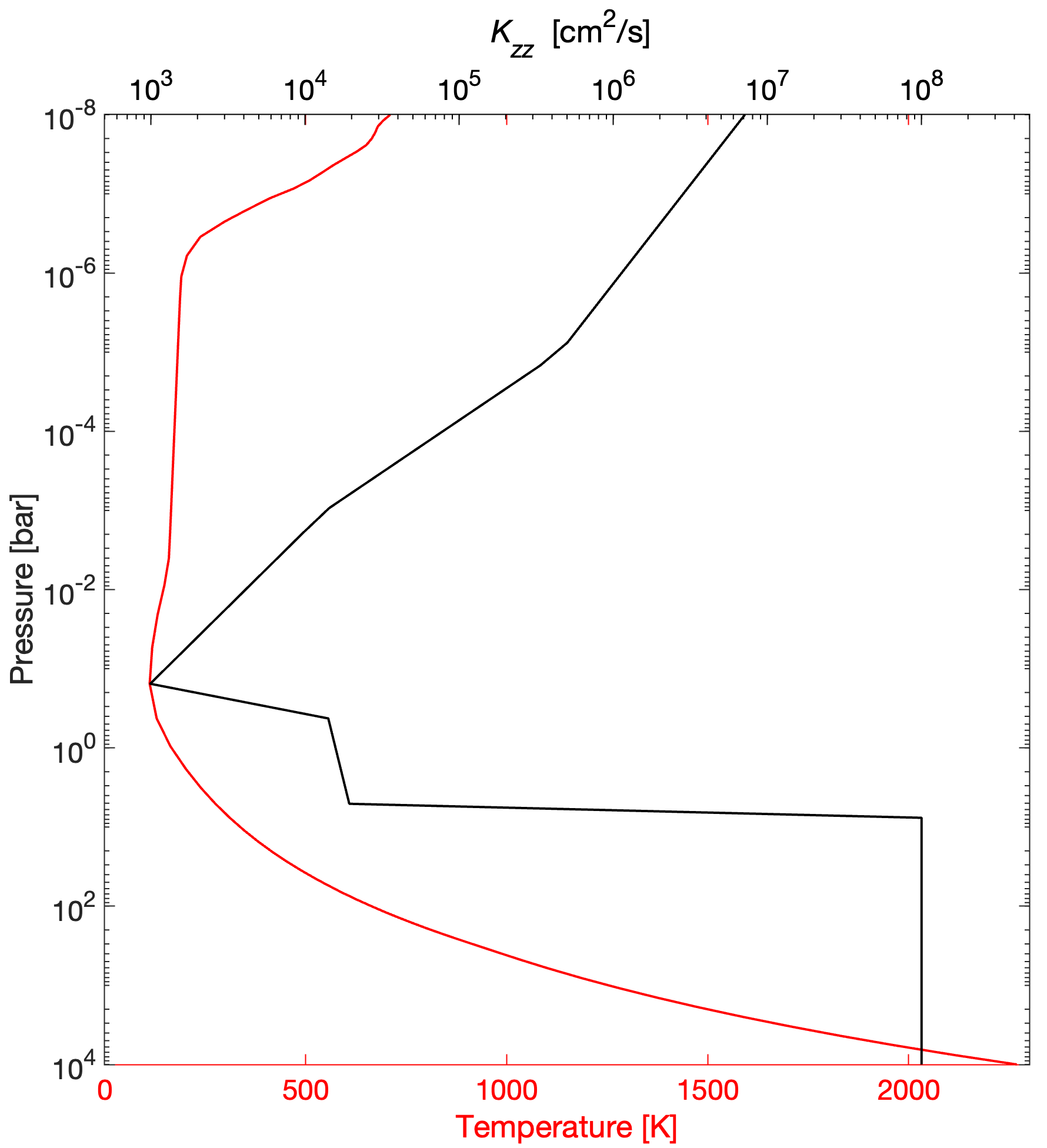}
     \caption{\footnotesize The temperature-pressure ($T$–$P$) and eddy diffusion coefficient ($K_{\text{zz}}$) profiles adopted in the current study. The red solid line represents the $T$–$P$ profile from \cite{seiff1998thermaljupiter, simon2006jupiter}, while the black solid line shows the $K_{\text{zz}}$ profile from \cite{moses2005photochemistry}. In this work, the deep atmospheric $K_{\text{zz}}$ value (uniform at pressures $P\gtrsim8$ bar) has been varied from $5\times10^6$ to $10^9$ cm$^2$/s, with a nominal value of $10^8$ cm$^2$/s shown in this figure.}
    \label{fig:tp_kzz_profiles}
\end{figure}

Figure A\ref{fig:tp_kzz_profiles} shows the $T$–$P$ and $K_{\text{zz}}$ profiles adopted in the current study.

%\FloatBarrier

\section{Chemical networks comparison} \label{sec:appendix_network_comaprisons}
\begin{figure}[htb!]
    \renewcommand{\figurename}{Figure C}
    %\centering    
    \includegraphics[width=0.45\textwidth]{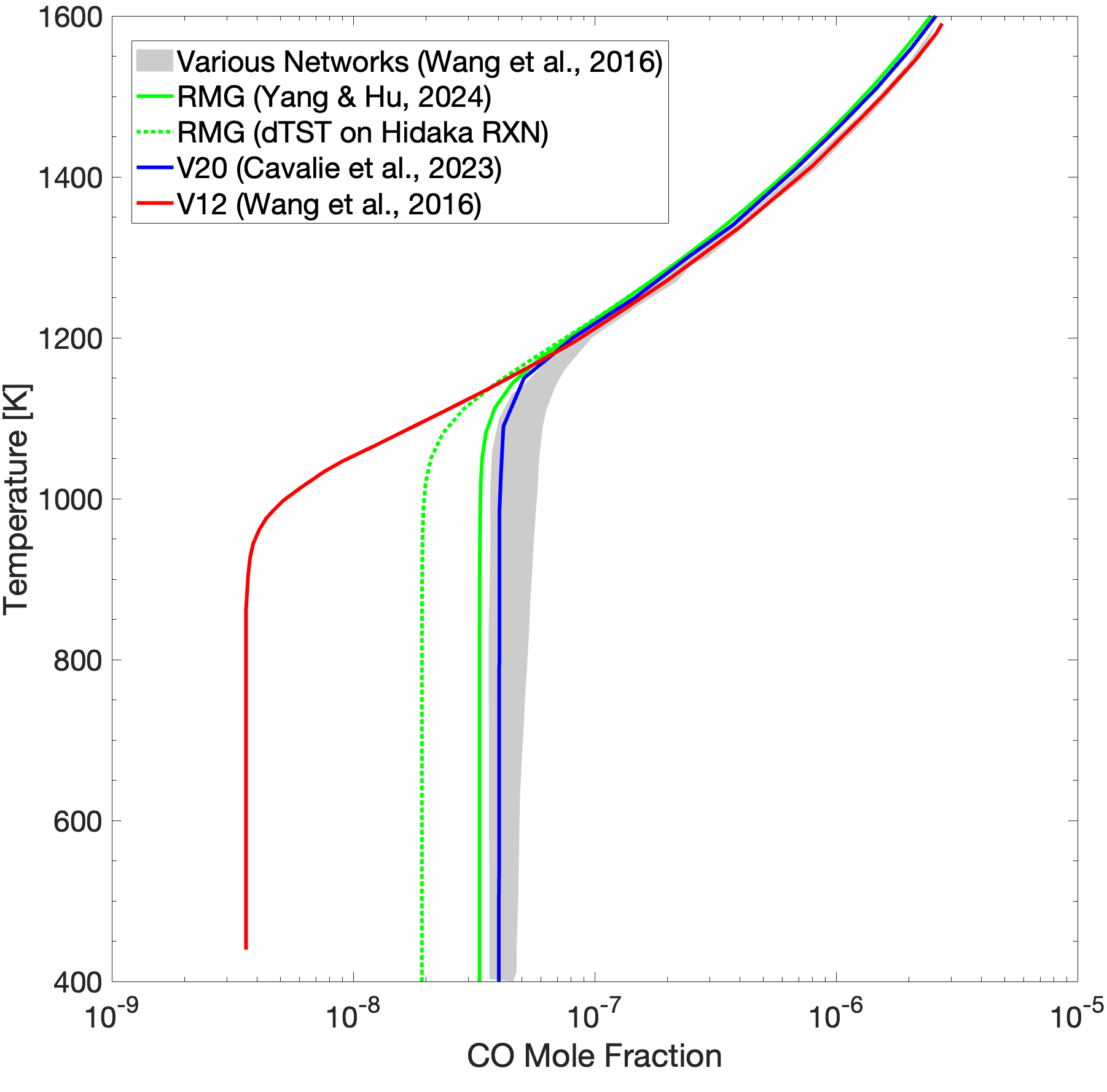}
     \caption{\footnotesize Vertical CO mole fraction profiles ($X_{\rm CO}$) in Jupiter’s troposphere were computed under identical conditions ($K_{\rm zz}=10^9$ [cm$^2$/s]; O/H = 7$\times Z_\odot$; He, C, N, S as in Section~\ref{subsec:elemental}), using various chemical networks. The RMG network from this study (green), V20 network \citep{venot2020new} (blue), and V12 network \citep{venot2012chemical} (red) show close agreement down to $\sim$1200 K, below which vertical mixing dominates. The green dotted line shows the RMG network with the Hidaka reaction rate updated from \citet{moses2011disequilibrium} to the dTST value from \cite{sanches2017novel}. The gray shaded region reflects the range of network predictions in \cite{Wang_2016_Jupiter}. While overall differences are modest, the RMG network predicts slightly more efficient CO-to-\ce{CH4} conversion than V20 and earlier networks under these oxygen-rich conditions.}
    \label{fig:chemical_network_comparison}
\end{figure}

Figure C~\ref{fig:chemical_network_comparison} shows comparisons among vertical mole fraction ($X_{\rm CO}$=[CO]/$n$, where $n$ is total number density of the atmosphere [molecules/cm$^3$]) profiles of CO in Jupiter's troposphere computed under identical conditions ($K_{\text{zz}}=10^9$ cm$^2$/s; He, C, N, and S abundances described in Section \ref{subsec:elemental}; O/H=7$\times Z_{\odot}$), as shown in Fig. 17 of \cite{Wang_2016_Jupiter}, but using various chemical networks.

%\FloatBarrier

\section{Chemical timescale fittings} \label{sec:appendix_chemicaltimescale}
\begin{figure}[hb!]
    \renewcommand{\figurename}{Figure D}
    %\centering    
    \includegraphics[width=0.47\textwidth]{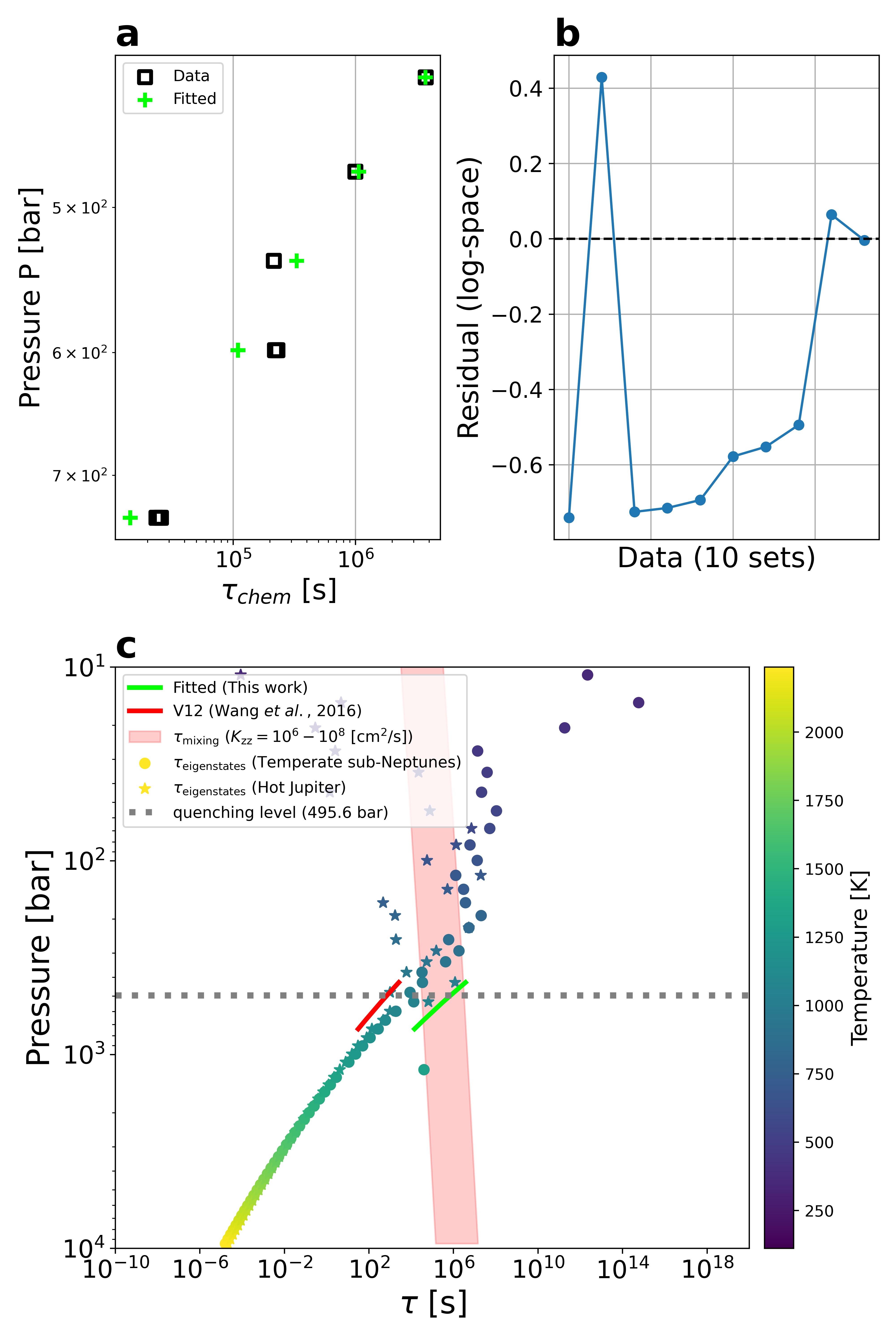}
     \caption{\footnotesize (a) Comparison between the thermochemistry-diffusion model derived data points of $\tau$\textsubscript{chem} described in Section \ref{subsec:t_chem_fitting} (black open squares) versus the fitted points (green cross symbols); (b) Residuals in logspace between the data points and fitted points; (c) Comparison among various chemical timescales of CO ($\tau$\textsubscript{chem}). The red solid line refers to the derived chemical timescale used in \cite{wang2015new}, The green solid line refers to the fitted chemical timescale in this work as described in Section \ref{subsec:t_chem_fitting}; The stars refer to the chemical timescale directly calculated under thermochemical equilibrium using the chemical network for the hot Jupiter chemistry used in \cite{yang2024automated}, while the circles refer to the chemical timescale directly calculated under thermochemical equilibrium using the chemical network for the temperate sub-Neptune chemistry used in \cite{yang2024chemical} (see Section~\ref{subsec:chemical_timescale_approach} and Section~\ref{sec:appendix_jacobian}). The chemical timescales derived from the two networks are nearly identical at temperatures above 1500 K. Note that the same chemical network is used for the green solid line and circles, but the green solid line was fitted using the method following \cite{wang2015new} whose method is described in Section~\ref{subsec:t_chem_fitting}, while the circles are directly calculated assuming thermochemical equilibrium using \texttt{Cantera} \citep{Goodwin_Cantera_An_Object-oriented_2024} and \texttt{scipy.linalg.eig} \citep{2020SciPy-NMeth} as described in Section~\ref{subsec:chemical_timescale_approach} and Section~\ref{sec:appendix_jacobian}.}
    \label{fig:timescalefitting}
\end{figure}

Table~D\ref{tab:dataset_timescale} summarizes the conditions for the 10 datasets used for the empirical fitting of the chemical timescale described in Section \ref{subsec:t_chem_fitting}. Figure~D\ref{fig:timescalefitting} presents the details of the fitting process and compares the resulting empirical timescale with that from the previous study by \cite{wang2015new}. Note that the chemical timescale derived by \cite{wang2015new} is based on the V12 chemical network developed by \cite{venot2012chemical}.

\begin{deluxetable}{cccc}[htb!]
\renewcommand{\tablename}{Table D}
\tablewidth{0.47\textwidth}
\tablecaption{Conditions for the 10 datasets used in the empirical fitting of the chemical timescale in this study\label{tab:dataset_timescale}}
\tablehead{
\colhead{O/H} & \colhead{\textit{K}\textsubscript{zz}[cm$^2$/s]} & \colhead{[CO]$^a$} & \colhead{$\tau$\textsubscript{CO} [s]}
}
\startdata
0.3$\times Z_{\odot}$$^b$ & $1\times10^8$ & $6.41\times10^{-10}$ & $2.30\times10^5$ \\
0.3$\times Z_{\odot}$ & $1\times10^9$ & $1.67\times10^{-9}$ & $2.56\times10^4$ \\
1$\times Z_{\odot}$ & $1\times10^8$ & $2.10\times10^{-9}$ & $2.15\times10^5$ \\
1.5$\times Z_{\odot}$ & $1\times10^8$ & $3.22\times10^{-9}$ & $2.26\times10^5$ \\
2.3$\times Z_{\odot}$ & $5\times10^6$ & $1.53\times10^{-9}$ & $3.75\times10^6$ \\
2.3$\times Z_{\odot}$ & $2\times10^7$ & $2.44\times10^{-9}$ & $9.96\times10^5$ \\
2.3$\times Z_{\odot}$ & $1\times10^8$ & $5.20\times10^{-9}$ & $2.24\times10^5$ \\
2.3$\times Z_{\odot}$ & $1\times10^9$ & $1.31\times10^{-8}$ & $2.50\times10^4$ \\
4$\times Z_{\odot}$ & $1\times10^8$ & $8.51\times10^{-9}$ & $2.19\times10^5$ \\
7$\times Z_{\odot}$ & $1\times10^9$ & $3.90\times10^{-8}$ & $2.35\times10^4$ \\
\enddata
\tablenotetext{a}{Quenched CO mixing ratio}
\tablenotetext{b}{Solar metallicity}
\end{deluxetable}

%\FloatBarrier

\section{Supplementary Materials for SNAP Simulations} \label{sec:appendix_SNAP}
\begin{figure}[hb!]
    \renewcommand{\figurename}{Figure E}
    \centering    
    \includegraphics[width=0.47\textwidth]{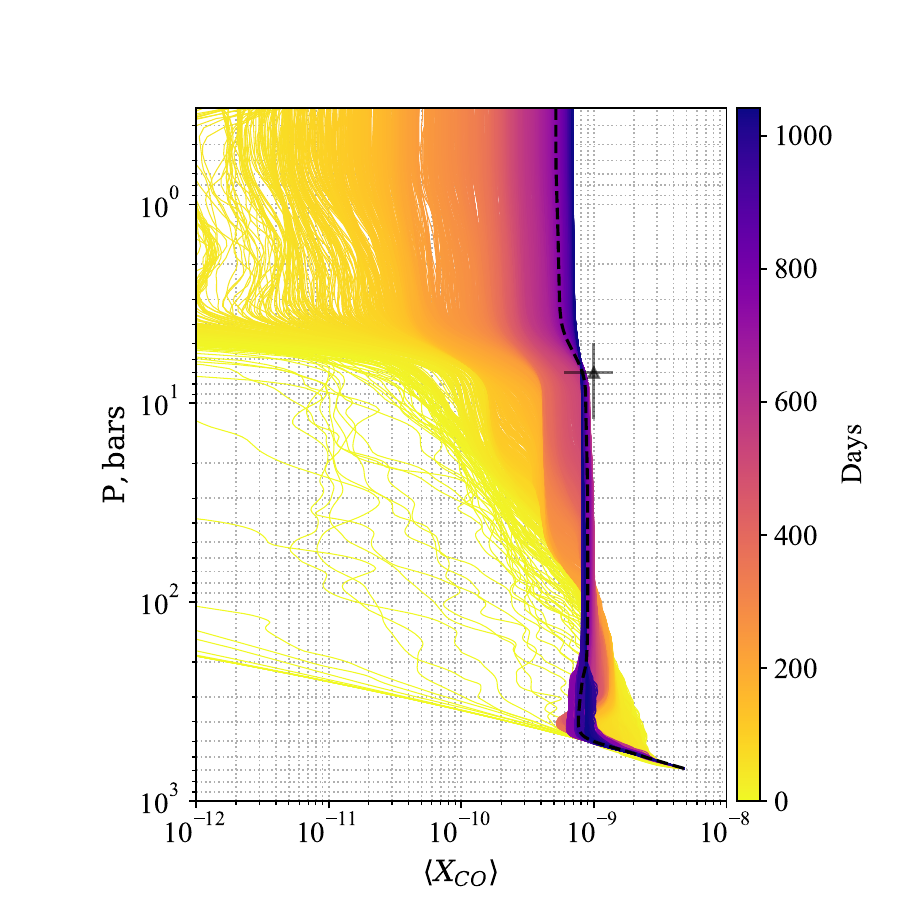}
     \caption{\footnotesize The time-dependent behavior of the horizontally averaged CO mole fraction for O/H=1.0$\times Z_{\odot}$. The solid gray triangle with error bars indicates the observed upper-tropospheric CO mole fraction from \citet{bezard2002carbon} and \citet{bjoraker2018gas}. Dashed line represents the time-averaged CO mole fraction for the O/H=1.0$\times Z_{\odot}$ scenario.}
    \label{fig:1xsolar}
\end{figure}

\begin{figure}[htb!]
    \renewcommand{\figurename}{Figure E}
    \centering    
    \includegraphics[width=0.49\textwidth]{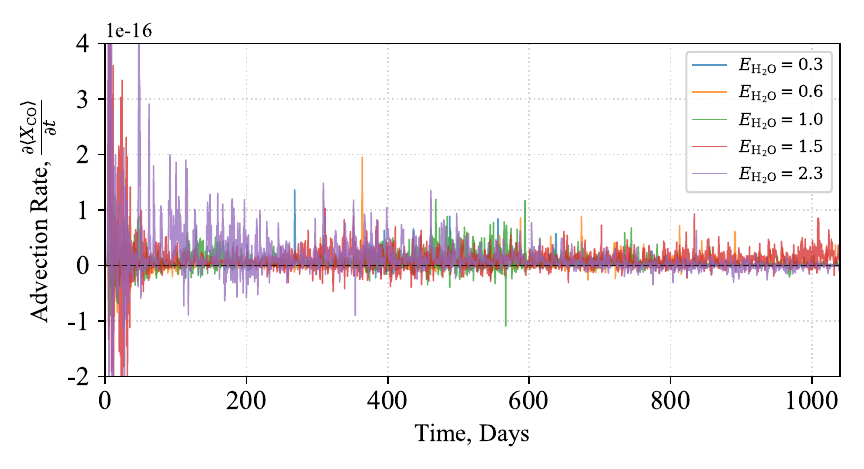}
     \caption{\footnotesize The mean tracer rate of change as the balance between hydrodynamical turbulent transport and chemical production and loss is achieved. By $\sim600$ days, the variation is consistently below machine precision ($10^{-16}$), suggestive of a chemical-diffusive steady-state.}
    \label{fig:f_prime_conv}
\end{figure}

Figure~E\ref{fig:1xsolar} shows the time-dependent behavior of the horizontally averaged CO mole fraction for O/H=1.0$\times Z_{\odot}$. Figure~E\ref{fig:f_prime_conv} shows the mean tracer rate of change as the balance between hydrodynamical turbulent transport and chemical production and loss, which is used to determine the convergence criteria of SNAP used in this work.

 \FloatBarrier

\section{ Chemical Timescale Approximation via Jacobian Eigenmodes Decomposition}
\label{sec:appendix_jacobian}
To assess the characteristic chemical timescales of species in Jupiter's deep atmosphere, we built the Jacobian matrix of the chemical system as follows.
First, we define the mole fraction column vector 
\begin{equation}
    \bm{X} = (X_1, X_2, ..., X_N)^T,
\end{equation}
where $X_{i}$ is the mole fraction of species $i$ at the given chemical system. $N$ is the total number of species in the chemical system of interest ($N$=89 for the chemical network built for the temperate sub-Neptune atmospheric simulation in \citet{yang2024chemical}, and $N$=106 for the chemical network built for the hot Jupiter atmospheric simulation in \citet{yang2024automated}, as shown in Figure~D\ref{fig:timescalefitting}). This mole fraction column vector $\bm{X}$ evolves according to
\begin{equation}
    \frac{d\bm{X}}{dt}=\bm{f}(\bm{X},T,P)
\end{equation}
at the given temperature $T$ [K] and pressure $P$ [Pa], where $\bm{f}$ denotes the vector of net production rates normalized by the total concentration, which we calculate using \texttt{Cantera}'s \texttt{Kinetics.net\_production\_rates} \citep{Goodwin_Cantera_An_Object-oriented_2024}. At around chemical equilibrium, denoted by $\bm{X}_{\rm eq}$ with $\bm{f}(\bm{X}_{\rm eq})=0$, we define a perturbation $\delta \bm{X}=\bm{X}-\bm{X}_{\rm eq}$. Assuming locally linearized dynamics around chemical equilibrium,
\begin{equation}
    \frac{d\bm{X}}{dt}=\frac{d}{dt}(\bm{X}_{\rm eq}+\delta\bm{X})=\frac{d (\delta \bm{X})}{dt}\sim\bm{J}\cdot\delta\bm{X}
\end{equation}
where the Jacobian matrix element is
\begin{equation}
    J_{ij}=\left.\frac{\partial f_i}{\partial X_j}\right|_{X = X_{\rm eq}}
\end{equation}
Numerically, we compute $J$ using central finite differences:
\begin{equation}
    J_{ij}\sim\frac{f_i(X_j+\Delta)-f_i(X_j-\Delta)}{2\Delta}
\end{equation}
where $\Delta$ is the perturbation of the mole fraction of species $j$ (i.e., $X_j$). We set this $\Delta=10^{-8}$ for CO analysis.
% We then build a reduced Jacobian $\bm{J_{\rm red}}$ by removing one species (CH$=$NH for temperate sub-Neptune chemical network and $^1$\ce{NH} for hot Jupiter chemical network). These species were added late in the mechanism generation and are therefore considered least important.
The characteristic modes of the system is given by the eigenvalue problem:
\begin{align}
    \bm{J}\cdot \bm{v}_k  &= \lambda_k\cdot \bm{v}_k, \\
    \bm{w}_k^{T}\cdot \bm{J} &= \lambda_k \cdot \bm{w}_k^{T}
\end{align}
where $\bm{v}_k$ and $\bm{w}_k^{T}$ are right and left eigenvectors, respectively, corresponding to $\lambda_k$. We used \texttt{scipy.linalg.eig} \citep{2020SciPy-NMeth} to compute these eigenvectors and corresponding eigenvalues of the system. Eigenvalues with negative real parts ($\Re(\lambda_k)<0$) represent decaying modes, with characteristic timescales
\begin{equation}
    \tau_k=-\frac{1}{\Re(\lambda_k)}
\end{equation}
To connect these modes to the chemical timescale of a specific species (CO in this study), we examined the left eigenvectors, which quantify how perturbations in each species project onto each mode. For a given target species index $k_{\rm CO}$, the row of the left eigenvector matrix provides the projection of CO participation onto every eigenmode. Among the decaying modes (i.e., $\Re(\lambda_k)<0$), we choose the $\Re(\lambda_k)$ in which CO has the largest magnitude of participation (i.e., the mode for which $\bm{w}_k^{T}$ has the maximum value). The results are shown in Figure~D\ref{fig:timescalefitting} for the two chemical networks (hot Jupiter chemical, shown with stars: temperate sub-Neptune, shown with circles) compared with empirical fittings from this work and from \citet{wang2015new}.

%% For this sample we use BibTeX plus aasjournalv7.bst to generate the
\bibliography{sample7}{}
\bibliographystyle{aasjournal}



\end{document}